\documentclass[english]{IEEEtran}
\usepackage[T1]{fontenc}
\usepackage[latin9]{inputenc}
\usepackage{color}
\usepackage{babel}
\usepackage{float}
\usepackage{amsmath}
\usepackage{amsthm}
\usepackage{amssymb}
\usepackage{graphicx}
\usepackage[unicode=true]
 {hyperref}

\makeatletter

\floatstyle{ruled}
\newfloat{algorithm}{tbp}{loa}
\providecommand{\algorithmname}{Algorithm}
\floatname{algorithm}{\protect\algorithmname}

\theoremstyle{plain}
\newtheorem{thm}{\protect\theoremname}
\theoremstyle{plain}
\newtheorem{lem}[thm]{\protect\lemmaname}
\theoremstyle{plain}
\newtheorem{cor}[thm]{\protect\corollaryname}
\theoremstyle{plain}
\newtheorem{prop}[thm]{\protect\propositionname}

\usepackage{algorithmic}
\usepackage{caption}

\newtheorem{assumption}{Assumption}

\makeatother

\providecommand{\corollaryname}{Corollary}
\providecommand{\lemmaname}{Lemma}
\providecommand{\propositionname}{Proposition}
\providecommand{\theoremname}{Theorem}

\begin{document}
\title{Solving High-Order Portfolios via Successive Convex Approximation
Algorithms}
\author{Rui~Zhou and Daniel~P.~Palomar,~\IEEEmembership{Fellow,~IEEE}\thanks{This work was supported by the Hong Kong RGC 16208917 research grant. }\thanks{The authors are with the Hong Kong University of Science and Technology
(HKUST), Clear Water Bay, Kowloon, Hong Kong (e-mail: \protect\href{mailto:rui.zhou@connect.ust.hk}{rui.zhou@connect.ust.hk};
\protect\href{mailto:palomar@ust.hk}{palomar@ust.hk}).}}
\maketitle
\begin{abstract}
The first moment and second central moments of the portfolio return,
a.k.a. mean and variance, have been widely employed to assess the
expected profit and risk of the portfolio. Investors pursue higher
mean and lower variance when designing the portfolios. The two moments
can well describe the distribution of the portfolio return when it
follows the Gaussian distribution. However, the real world distribution
of assets return is usually asymmetric and heavy-tailed, which is
far from being a Gaussian distribution. The asymmetry and the heavy-tailedness
are characterized by the third and fourth central moments, i.e., skewness
and kurtosis, respectively. Higher skewness and lower kurtosis are
preferred to reduce the probability of extreme losses. However, incorporating
high-order moments in the portfolio design is very difficult due to
their non-convexity and rapidly increasing computational cost with
the dimension. In this paper, we propose a very efficient and convergence-provable
algorithm framework based on the successive convex approximation (SCA)
algorithm to solve high-order portfolios. The efficiency of the proposed
algorithm framework is demonstrated by the numerical experiments.
\end{abstract}


\section{\textcolor{black}{Introduction}}

Modern portfolio theory has developed rapidly since Harry Markowitz\textquoteright s
seminal paper in 1952, which proposed the mean-variance framework
to pursue the trade-off between maximizing the portfolio's profit
and minimizing the risk \cite{markowitz1952eltit}. The profit and
risk of a portfolio are measured by the mean and variance, i.e., the
first moment and the second central moments, of the portfolio return.
The mean-variance framework assumes that the investors prefer a quadratic
utility or that the returns of assets follow a Gaussian distribution
\cite{kolm201460}.

However, the mean-variance framework is not widely used in the real
market investment. One of the main reasons is that returns of assets
in real markets are seldom Gaussian distributed. They are usually
asymmetric and more likely to contain outliers or exhibit a heavier
tail, making the portfolio return also asymmetric and heavy-tailed
\cite{adcock2015skewed,resnick2007heavy}. Meanwhile, most investors
would be willing to accept lower expected profit and higher volatility
in exchange for more positively skewed and less heavy-tailed portfolio
return \cite{harvey2000conditional,jobst2001tail,ang2006downside}.
This aspiration has been beyond the characterization of the mean-variance
framework. Apart from that, the investors might have different tastes
in utility functions. Sometimes the shapes of these utility functions
can be significantly different from the quadratic one. 

To make up the drawbacks of the mean-variance framework, we need to
take high-order moments of the portfolio return into consideration.
The asymmetry and heavy-tailedness of portfolio return are well captured
by its third and fourth central moments, i.e., skewness and kurtosis.
A higher skewness usually means that the portfolio return admits a
more positively skewed shape, while the lower kurtosis usually corresponds
to thinner tail. We can extend the mean-variance framework by directly
incorporating the high-order moments to obtain the mean-variance-skewness-kurtosis
(MVSK) framework, where we shall try to strike a balance between maximizing
the mean and skewness (odd moments) while minimizing the variance
and kurtosis (even moments) \cite{dinh2011efficient,boudt2015higher,kshatriya2018genetic}.
Besides, such extension can be seen as approximating a general expected
utility function with its Taylor series expansion truncated to the
four most important order terms \cite{jean1971extension}. There also
exist some other high-order portfolios within the MVSK framework.
For example, the MVSK tilting portfolios \cite{BOUDT2020e03516} are
obtained by ``tilting'' a given portfolio to the MVSK efficient
frontier.

Although there are many advantages of the MVSK framework, solving
such high-order portfolio optimization problems is quite challenging.\textcolor{black}{{}
First, the third and fourth central moments are both nonconvex functions,
making the problems in general NP-hard \cite{murty1985some}.} These
problems are traditionally solved by some \textcolor{black}{metaheuristic}
optimization tools, e.g., differential evolution \cite{maringer2009global}
and genetic algorithms \cite{kshatriya2018genetic}.\textcolor{black}{{}
However, they are essentially performing a time-consuming random search
 \cite{blum2003metaheuristics,savine2018modern}.} A method based
on the Difference of Convex (DC) algorithm was proposed to solve the
MVSK portfolio problem to a stationary point \cite{dinh2011efficient},
but it converges too slowly and that it is only applicable to small-size
problems. Second, the complexity of computing the value or the gradients
of high-order moments grows rapidly with the problem dimension. The
classical general gradient descent method and backtracking line search
also become inapplicable when the problem dimension grows large. Therefore,
it is meaningful and necessary to design efficient algorithms for
solving high-order portfolios.

To this end, the major goal of this paper is to develop an efficient
algorithm framework based on the successive convex approximation (SCA)
to solve high-order portfolios. The SCA algorithm solves the original
intractable problem by constructing and solving a sequence of strongly
convex approximating problems \cite{sun2016majorization,nedic2018multi,facchinei2020ghost}.
In this paper, we propose an easy approach to construct the  approximation
for the nonconvex functions. This allows to construct a sequence of
convex problems compatible with existing efficient solvers that can
obtain the solutions to the original high-order portfolio optimization
problems. The convergence of the proposed algorithm framework to a
stationary point is established. In addition, owing to their low computational
complexity, the algorithms are amenable for high-dimensional applications.
Extensive numerical experiments are performed to corroborate our claims.

The paper is organized as follows. We first give the preliminary knowledge
on the high-order moments of portfolio return in Section \ref{sec: preliminaries}
and then pose the problem formulations in Section \ref{sec: problem formulation}.
The SCA algorithm and its special cases are introduced in Section
\ref{sec: SCA}. In Section \ref{sec: solving MVSK} and Section \ref{sec: solving MVSK tilting},
we derive our algorithms based on the SCA algorithm to solve the high-order
portfolios. The complexity and convergence analysis of the proposed
algorithms are discussed in Section \ref{sec: compl and convg anal}.
In Section \ref{sec: other formulations}, we present some other formulations
of high-order portfolio problems and indicate the applicability of
our proposed algorithm framework. The numerical experiments are given
in Section \ref{sec: Numerical Experiments}. Finally, the conclusion
of this paper is summarized in Section \ref{sec: conclusion}.

\section{Preliminaries: the Moments of Portfolio Return \label{sec: preliminaries}}

Denote by $\mathbf{r}\in\mathbb{R}^{N}$ the returns of $N$ assets
and $\mathbf{w}\in\mathbb{R}^{N}$ the portfolio weights. The return
of this portfolio is $\mathbf{w}^{T}\mathbf{r}$ with expected value,
i.e., the first moment
\begin{equation}
\phi_{1}\left(\mathbf{w}\right)=\text{E}\left[\mathbf{w}^{T}\mathbf{r}\right]=\mathbf{w}^{T}\boldsymbol{\mu},
\end{equation}
where $\boldsymbol{\mu}=\text{E}\left(\mathbf{r}\right)$ is the mean
vector of the assets' returns. Denote by $\tilde{\mathbf{r}}=\mathbf{r}-\boldsymbol{\mu}$
the centered returns, the $q$-th central moment of the portfolio
return is $\text{E}\left[\left(\mathbf{w}^{T}\mathbf{r}-\mathbf{w}^{T}\boldsymbol{\mu}\right)^{q}\right]=\text{E}\left[\left(\mathbf{w}^{T}\tilde{\mathbf{r}}\right)^{q}\right]$,
which gives us the following:
\begin{itemize}
\item The second central moment, a.k.a. variance, of the portfolio return
is
\begin{equation}
\begin{aligned}\phi_{2}\left(\mathbf{w}\right) & =\text{E}\left[\left(\mathbf{w}^{T}\tilde{\mathbf{r}}\right)^{2}\right]\\
 & =\text{E}\left[\mathbf{w}^{T}\tilde{\mathbf{r}}\tilde{\mathbf{r}}^{T}\mathbf{w}\right]\\
 & =\mathbf{w}^{T}\boldsymbol{\Sigma}\mathbf{w},
\end{aligned}
\end{equation}
where $\boldsymbol{\Sigma}=\text{E}\left[\tilde{\mathbf{r}}\tilde{\mathbf{r}}^{T}\right]$
is the covariance matrix.
\item The third central moment, a.k.a. skewness, of the portfolio return
is
\begin{equation}
\begin{aligned}\phi_{3}\left(\mathbf{w}\right) & =\text{E}\left[\left(\mathbf{w}^{T}\tilde{\mathbf{r}}\right)^{3}\right]\\
 & =\text{E}\left[\mathbf{w}^{T}\tilde{\mathbf{r}}\tilde{\mathbf{r}}^{T}\mathbf{w}\tilde{\mathbf{r}}^{T}\mathbf{w}\right]\\
 & =\text{E}\left[\mathbf{w}^{T}\tilde{\mathbf{r}}\left(\tilde{\mathbf{r}}^{T}\otimes\tilde{\mathbf{r}}^{T}\right)\left(\mathbf{w}\otimes\mathbf{w}\right)\right]\\
 & =\mathbf{w}^{T}\boldsymbol{\Phi}\left(\mathbf{w}\otimes\mathbf{w}\right),
\end{aligned}
\end{equation}
where $\boldsymbol{\Phi}=\text{E}\left[\tilde{\mathbf{r}}\left(\tilde{\mathbf{r}}^{T}\otimes\tilde{\mathbf{r}}^{T}\right)\right]$
is the co-skewness matrix. 
\item The fourth central moment, a.k.a. kurtosis, of the portfolio return
is
\begin{equation}
\begin{aligned}\phi_{4}\left(\mathbf{w}\right) & =\text{E}\left[\left(\mathbf{w}^{T}\tilde{\mathbf{r}}\right)^{4}\right]\\
 & =\text{E}\left[\mathbf{w}^{T}\tilde{\mathbf{r}}\tilde{\mathbf{r}}^{T}\mathbf{w}\tilde{\mathbf{r}}^{T}\mathbf{w}\tilde{\mathbf{r}}^{T}\mathbf{w}\right]\\
 & =\text{E}\left[\mathbf{w}^{T}\tilde{\mathbf{r}}\left(\tilde{\mathbf{r}}^{T}\otimes\tilde{\mathbf{r}}^{T}\right)\left(\mathbf{w}\otimes\mathbf{w}\right)\tilde{\mathbf{r}}^{T}\mathbf{w}\right]\\
 & =\text{E}\left[\mathbf{w}^{T}\tilde{\mathbf{r}}\left(\tilde{\mathbf{r}}^{T}\otimes\tilde{\mathbf{r}}^{T}\otimes\tilde{\mathbf{r}}^{T}\right)\left(\mathbf{w}\otimes\mathbf{w}\otimes\mathbf{w}\right)\right]\\
 & =\mathbf{w}^{T}\boldsymbol{\Psi}\left(\mathbf{w}\otimes\mathbf{w}\otimes\mathbf{w}\right),
\end{aligned}
\end{equation}
where $\boldsymbol{\Psi}=\text{E}\left[\tilde{\mathbf{r}}\left(\tilde{\mathbf{r}}^{T}\otimes\tilde{\mathbf{r}}^{T}\otimes\tilde{\mathbf{r}}^{T}\right)\right]$
is the co-kurtosis matrix.
\end{itemize}
The gradients of $\phi_{1}\left(\mathbf{w}\right)$ and $\phi_{2}\left(\mathbf{w}\right)$
w.r.t. $\mathbf{w}$ are $\boldsymbol{\mu}$ and $2\boldsymbol{\Sigma}\mathbf{w}$,
while their Hessians are $\mathbf{0}$ and $2\boldsymbol{\Sigma}$,
respectively. But the gradient and the Hessian of $\phi_{3}\left(\mathbf{w}\right)$
and $\phi_{4}\left(\mathbf{w}\right)$ are more complicated to derive
and we give the next some useful results.
\begin{lem}
\label{lem: gradient and hessian of moments}The gradient and Hessian
of the skewness and kurtosis are given by:
\begin{equation}
\begin{aligned}\triangledown\phi_{3}\left(\mathbf{w}\right) & =3\boldsymbol{\Phi}\left(\mathbf{w}\otimes\mathbf{w}\right),\\
\triangledown\phi_{4}\left(\mathbf{w}\right) & =4\boldsymbol{\Psi}\left(\mathbf{w}\otimes\mathbf{w}\otimes\mathbf{w}\right),\\
\triangledown^{2}\phi_{3}\left(\mathbf{w}\right) & =6\boldsymbol{\Phi}\left(\mathbf{I}\otimes\mathbf{w}\right),\\
\triangledown^{2}\phi_{4}\left(\mathbf{w}\right) & =12\boldsymbol{\Psi}\left(\mathbf{I}\otimes\mathbf{w}\otimes\mathbf{w}\right).
\end{aligned}
\end{equation}
\end{lem}
\begin{IEEEproof}
See Appendix \ref{apx: proof grad and hsn}.
\end{IEEEproof}
\begin{cor}
\label{cor: grad hsn relations}The gradient and Hessian of the skewness
and kurtosis admit the following relations:
\begin{equation}
\triangledown\phi_{3}\left(\mathbf{w}\right)=\frac{1}{2}\triangledown^{2}\phi_{3}\left(\mathbf{w}\right)\mathbf{w},\label{eq: skewness grad hsn relations}
\end{equation}
\begin{equation}
\triangledown\phi_{4}\left(\mathbf{w}\right)=\frac{1}{3}\triangledown^{2}\phi_{4}\left(\mathbf{w}\right)\mathbf{w}.\label{eq: kurtosis grad hsn relations}
\end{equation}
\end{cor}
\begin{IEEEproof}
Using Lemma \ref{lem: gradient and hessian of moments}, we have $3\phi_{3}\left(\mathbf{w}\right)=\mathbf{w}^{T}\triangledown\phi_{3}\left(\mathbf{w}\right)$.
Then taking the derivative of both sides w.r.t. $\mathbf{w}$, we
get $3\triangledown\phi_{3}\left(\mathbf{w}\right)=\triangledown\phi_{3}\left(\mathbf{w}\right)+\triangledown^{2}\phi_{3}\left(\mathbf{w}\right)\mathbf{w}$,
which further derives equation (\ref{eq: skewness grad hsn relations}).
Equation (\ref{eq: kurtosis grad hsn relations}) can be derived similarly. 
\end{IEEEproof}
\textcolor{black}{Note that $\triangledown^{2}\phi_{3}\left(\mathbf{w}\right)=6\sum_{k=1}^{N}\Phi_{ij}^{(k)}w_{k}$
and $\triangledown^{2}\phi_{4}\left(\mathbf{w}\right)=12\sum_{k,l=1}^{N}\Psi_{ij}^{(k,l)}w_{k}w_{l}$
can be easily obtained from Lemma \ref{lem: gradient and hessian of moments},
where $\Phi_{ij}^{(k)}=\text{E}\left[\tilde{r}_{i}\tilde{r}_{j}\tilde{r}_{k}\right]$
and $\Psi_{ij}^{(k,l)}=\text{E}\left[\tilde{r}_{i}\tilde{r}_{j}\tilde{r}_{k}\tilde{r}_{l}\right]$
are the corresponding elements of matrices $\boldsymbol{\Phi}$ and
$\boldsymbol{\Psi}$. }

\begin{figure}
\begin{centering}
\includegraphics[scale=0.47]{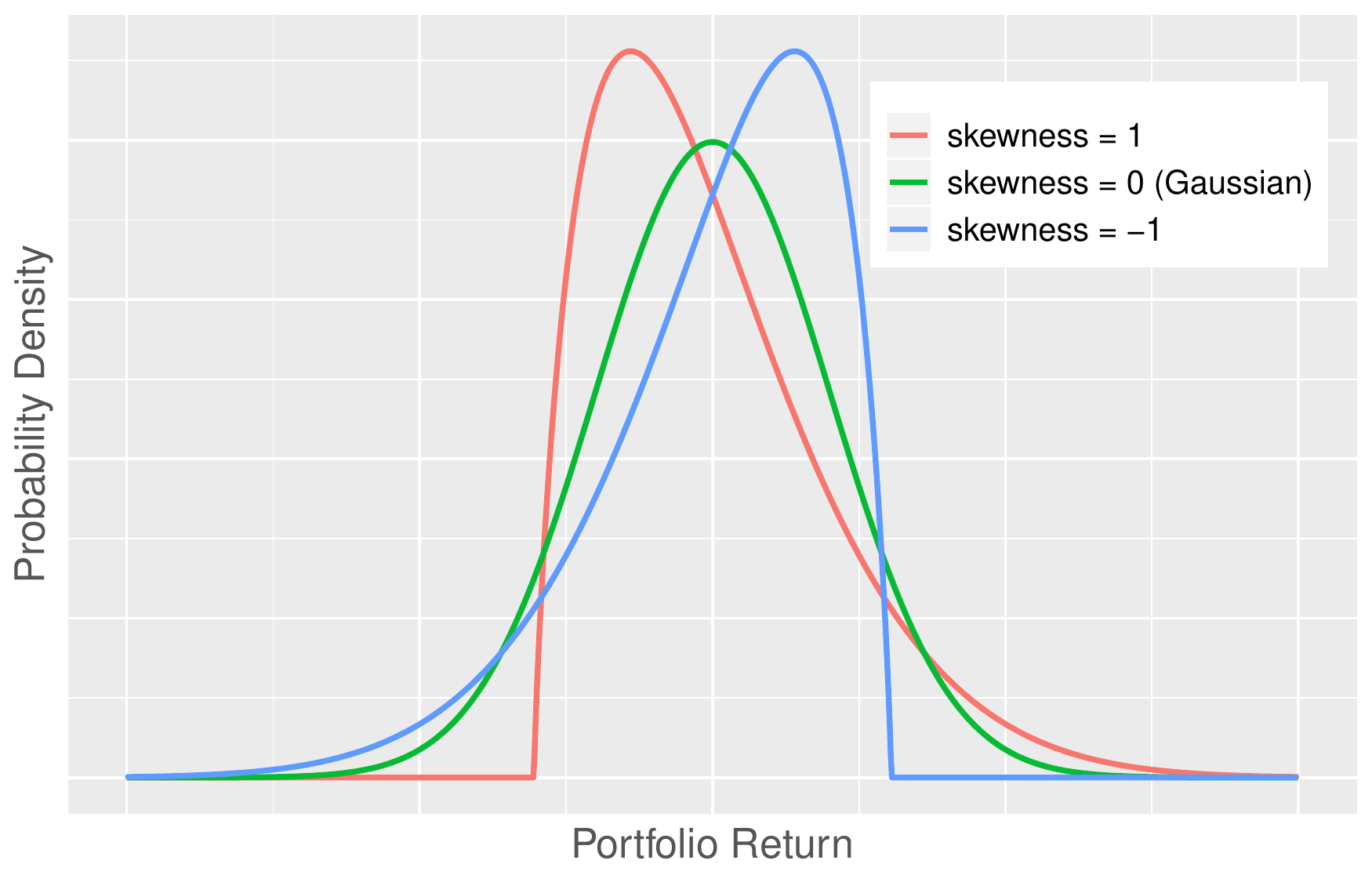}
\par\end{centering}
\caption{The implication of skewness. \label{fig: skewness illustrate}}

\end{figure}

\begin{figure}
\begin{centering}
\includegraphics[scale=0.47]{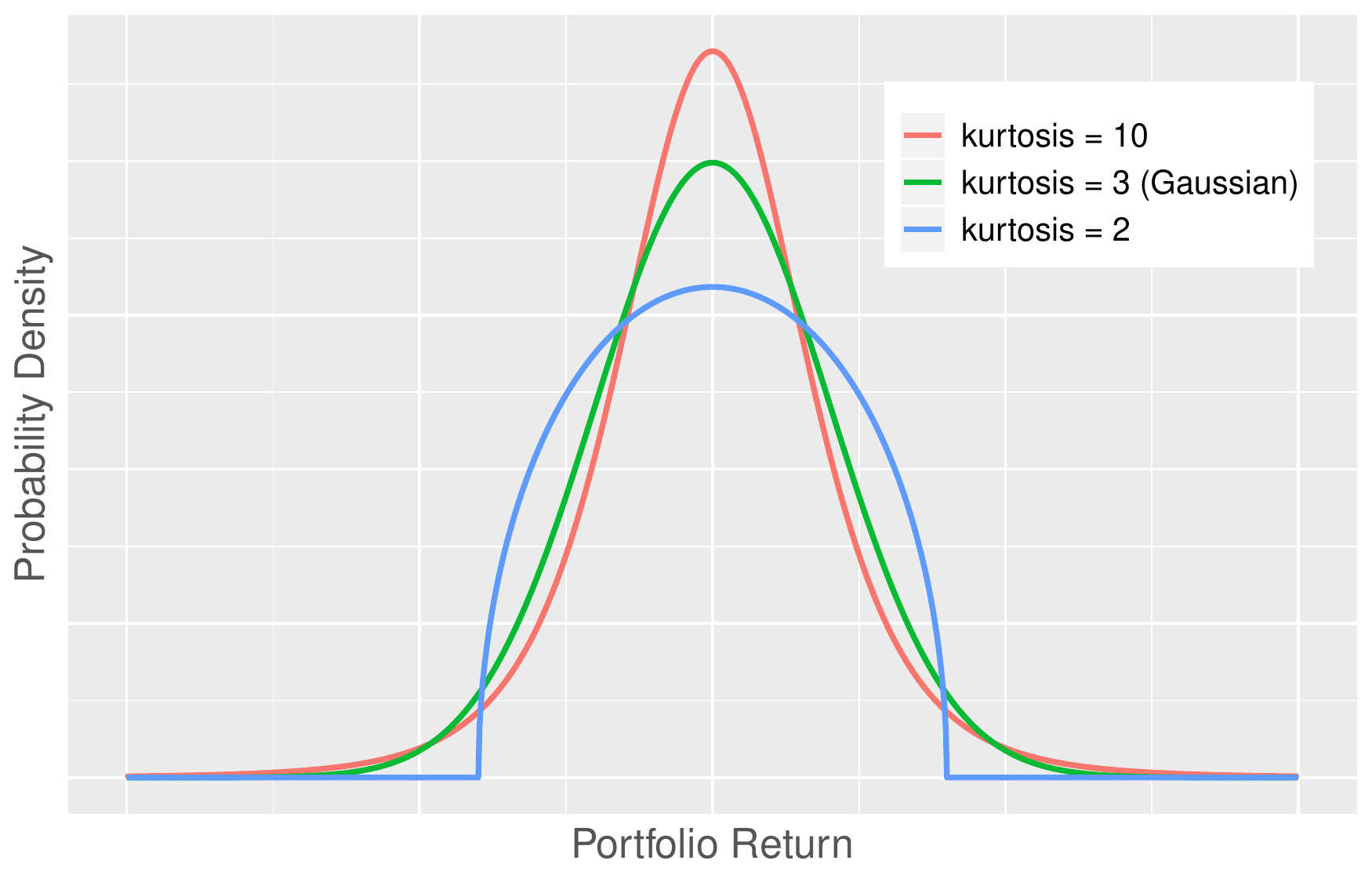}
\par\end{centering}
\caption{The implication of kurtosis. \label{fig: kurtosis illustrate}}
\end{figure}

A high expected value and low variance of the portfolio return are
naturally chased by investors to increase the profit and decrease
the risk. Besides, in the non-Gaussian case, a high skewness and low
kurtosis are also desirable as they can reduce the probability of
extreme losses. As shown in Figure \ref{fig: skewness illustrate},
a positively skewed portfolio return is significantly less likely
to suffer extreme losses than a negatively skewed one. Besides, we
can see from Figure \ref{fig: kurtosis illustrate} that a lower kurtosis
shows also a thinner tail, which alleviates the appearance of extreme
returns. In general, investors have a preference for odd moments while
dislike even moments.


\section{Problem Formulation \label{sec: problem formulation}}

\subsection{MVSK Portfolio}

The classical Markowitz's mean-variance (MV) portfolio \cite{markowitz1952eltit}
is obtained by solving the following problem:
\begin{equation}
\begin{aligned}\underset{\mathbf{w}}{\mathsf{minimize}}\,\,\, & \,\,\,\,-\mathbf{w}^{T}\boldsymbol{\mu}+\lambda\mathbf{w}^{T}\boldsymbol{\Sigma}\mathbf{w}\\
\mathsf{subject\,\,to} & \quad\mathbf{w}\in\mathcal{W},
\end{aligned}
\label{eq: mean-variance problem}
\end{equation}
where $\lambda\ge0$ is a parameter striking a balance between the
expected return ($\mathbf{w}^{T}\boldsymbol{\mu}$) and the portfolio
risk (defined by the variance $\mathbf{w}^{T}\boldsymbol{\Sigma}\mathbf{w}$),
$\mathcal{W}$ is the feasible set of portfolio weights, which we
set as
\begin{equation}
\mathcal{W}=\left\{ \mathbf{w}\vert\mathbf{1}^{T}\mathbf{w}=1,\Vert\mathbf{w}\Vert_{1}\le L\right\} ,\label{eq: constrains for w}
\end{equation}
 where $L\ge1$ is the leverage constraint of the portfolio \cite{zhao2019optimal}.
Specifically, when $L=1$, $\mathcal{W}$ reduces to the no shorting
constraint: $\left\{ \mathbf{w}\vert\mathbf{1}^{T}\mathbf{w}=1,\mathbf{w}\ge\mathbf{0}\right\} $.
The expected mean and the expected variance are actually the first
moment and the second central moment of the portfolio return. However,
the real world assets return usually appears to be asymmetric and
of extreme values, which is beyond the characterization of first two
moments. It is reasonable to consider the third and fourth central
moments in the portfolio design. A natural way to incorporate the
two higher-order moments is revising the objective of problem (\ref{eq: mean-variance problem})
to achieve the mean-variance-skewness-kurtosis portfolio design problem
\cite{dinh2011efficient,boudt2015higher,kshatriya2018genetic}:
\begin{equation}
\begin{aligned}\underset{\mathbf{w}}{\mathsf{minimize}}\,\, & \,\,\,\,\,f\left(\mathbf{w}\right)=-\lambda_{1}\phi_{1}\left(\mathbf{w}\right)+\lambda_{2}\phi_{2}\left(\mathbf{w}\right)\\
 & \hspace{1.6cm}-\lambda_{3}\phi_{3}\left(\mathbf{w}\right)+\lambda_{4}\phi_{4}\left(\mathbf{w}\right)\\
\mathsf{subject\,\,to} & \quad\mathbf{w}\in\mathcal{W},
\end{aligned}
\label{eq: mvsk portfolio}
\end{equation}
where  $\lambda_{1},\lambda_{2},\lambda_{3},\lambda_{4}\ge0$ are
the parameters for combining the four moments of the portfolio return.

\subsection{MVSK Tilting Portfolio}

Directly solving the problem (\ref{eq: mvsk portfolio}) leads us
to the MVSK efficient frontier, where we cannot improve any moment
without impairing other moments. However, the investors might want
to modify another existing portfolio $\mathbf{w}_{0}$ toward a MVSK
efficient portfolio. This can be done by tilting these portfolios
in a direction that increases their first moment and third central
moment and decreases their second and fourth central moments \cite{BOUDT2020e03516},
i.e., \textcolor{black}{
\begin{equation}
\begin{aligned}\underset{\mathbf{w},\delta}{\mathsf{maximize}}\,\, & \quad\delta\\
\mathsf{subject\,\,to} & \quad\phi_{1}\left(\mathbf{w}\right)\ge\phi_{1}\left(\mathbf{w}_{0}\right)+d_{1}\delta,\\
 & \quad\phi_{2}\left(\mathbf{w}\right)\le\phi_{2}\left(\mathbf{w}_{0}\right)-d_{2}\delta,\\
 & \quad\phi_{3}\left(\mathbf{w}\right)\ge\phi_{3}\left(\mathbf{w}_{0}\right)+d_{3}\delta,\\
 & \quad\phi_{4}\left(\mathbf{w}\right)\le\phi_{4}\left(\mathbf{w}_{0}\right)-d_{4}\delta,\\
 & \quad\left(\mathbf{w}-\mathbf{w}_{0}\right)^{T}\boldsymbol{\Sigma}\left(\mathbf{w}-\mathbf{w}_{0}\right)\le\kappa^{2},\\
 & \quad\mathbf{w}\in\mathcal{W},\delta\ge0,
\end{aligned}
\label{eq: mvsk tilting problem}
\end{equation}
where $\mathbf{d}=\left[d_{1},d_{2},d_{3},d_{4}\right]\ge\mathbf{0}$
is the tilting direction,} $\phi_{i}\left(\mathbf{w}_{0}\right),i=1,2,3,4$
are the moments of $\mathbf{w}_{0}$ (starting point) for tilting,
$\kappa^{2}$ determines the maximum tracking error volatility of
$\mathbf{w}$ with respect to the reference portfolio $\mathbf{w}_{0}$.

\subsection{Difficulty of Solving High-Order Portfolios}

The MVSK portfolio optimization problem (\ref{eq: mvsk portfolio})
and MVSK tilting portfolio optimization problems (\ref{eq: mvsk tilting problem})
are very difficult to solve for two reasons:
\begin{enumerate}
\item \textbf{Non-convexity}: the third and fourth central moments, i.e.,
$\phi_{3}\left(\mathbf{w}\right)$ and $\phi_{4}\left(\mathbf{w}\right)$,
are non-convex on $\mathbf{w}$, making the problem (\ref{eq: mvsk portfolio})
and the problem (\ref{eq: mvsk tilting problem}) both non-convex
problems. 
\item \textbf{Computational complexity}: $\boldsymbol{\Psi}$ is of dimension
$N\times N^{3}$, which means the memory complexity is $\mathcal{O}\left(N^{4}\right)$
and the computational complexity of one single evaluation of the fourth
moment is $\mathcal{O}\left(N^{4}\right)$. Lemma \ref{lem: gradient and hessian of moments}
shows that the computational complexity for computing the gradient
of the fourth central moment is also $\mathcal{O}\left(N^{4}\right)$.
Then the general gradient descent method and backtracking line search
are inappropriate to the high-order portfolio problem.
\end{enumerate}
Due to the non-convexity, the classical convex optimization methods
are not applicable, while the general gradient method is also not
applicable due to the expensive cost of gradient computation. It is
necessary to design a specific algorithm to efficiently solve high-order
portfolios. Such an algorithm should converge fast and avoid evaluating
the gradients or value of high-order moments frequently. This paper
proposes a very efficient algorithm framework to solve the high-order
portfolio optimization problem based on the SCA algorithm. But before
that, some background on the SCA algorithm is due in the next section.

\section{The Successive Convex Approximation Algorithm \label{sec: SCA}}

The successive convex approximation (SCA) algorithm is a general framework
especially designed for solving non-convex optimization problems.
Instead of solving the original intractable optimization problem,
it resorts to successively solving a sequence of strongly convex approximating
problems. The convergence of the SCA algorithm can be guaranteed under
mild assumptions.

Specifically, consider a nonconvex constrained optimization problem,
\begin{equation}
\begin{aligned}\underset{\mathbf{x}}{\mathsf{minimize}}\,\,\, & \quad f\left(\mathbf{x}\right)\\
\mathsf{subject\,\,to} & \quad g_{i}\left(\mathbf{x}\right)\le0,\,\,i=1,\dots,m,\\
 & \quad\mathbf{x}\in\mathcal{K},
\end{aligned}
\label{eq: SCA original problem}
\end{equation}
where $f\left(\mathbf{x}\right)$ and $g_{i}\left(\mathbf{x}\right)$
are nonconvex functions and $\mathcal{K}$ is a convex set. In order
to solve the problem (\ref{eq: SCA original problem}), which is directly
intractable, we may turn to successively solving a sequence of strongly
convex approximating problems. Denote by $\mathbf{x}^{k}$ the current
iterate at $k$-th iteration, then the SCA algorithm constructs a
strongly convex approximating problem for (\ref{eq: SCA original problem})
as \cite{facchinei2020ghost}:
\begin{equation}
\begin{aligned}\underset{\mathbf{x}}{\mathsf{minimize}}\,\,\, & \quad\tilde{f}\left(\mathbf{x};\mathbf{x}^{k}\right)\\
\mathsf{subject\,\,to} & \quad\tilde{g}_{i}\left(\mathbf{x};\mathbf{x}^{k}\right)\le\eta\left(\mathbf{x}^{k}\right),i=1,\dots,m,\\
 & \quad\Vert\mathbf{x}-\mathbf{x}^{k}\Vert_{\infty}\le\beta,\\
 & \quad\mathbf{x}\in\mathcal{K},
\end{aligned}
\label{eq: SCA surrogate problem}
\end{equation}
where $\tilde{f}\left(\mathbf{x};\mathbf{x}^{k}\right)$ and $\tilde{g}_{i}\left(\mathbf{x};\mathbf{x}^{k}\right)$
are the approximating functions for $f\left(\mathbf{x}\right)$ and
$g_{i}\left(\mathbf{x}\right)$ at $\mathbf{x}^{k}$, the quantity
$\eta\left(\mathbf{x}^{k}\right)$ in the surrogate constraints serves
to suitably enlarge the feasible set of the subproblem to ensure it
is always nonempty, and $\beta$ is a user-chosen positive constant.
The term $\eta\left(\mathbf{x}^{k}\right)$ is defined as
\begin{equation}
\begin{aligned}\eta\left(\mathbf{x}^{k}\right) & \triangleq\left(1-\theta\right)\max_{i}\left\{ g_{i}(\mathbf{x}^{k})_{+}\right\} \\
 & \quad+\theta\min_{\mathbf{x}}\left\{ \max_{i}\big\{\tilde{g}_{i}(\mathbf{x};\mathbf{x}^{k})_{+}\big\}\big|\mathbf{x}\in\mathcal{K}\right\} ,
\end{aligned}
\end{equation}
with $\theta\in\left(0,1\right)$. The general SCA algorithm generates
the sequence $\left\{ \mathbf{x}^{k}\right\} $ as
\begin{equation}
\begin{cases}
\hat{\mathbf{x}}^{k+1} & \leftarrow\text{solve the problem \eqref{eq: SCA surrogate problem}},\\
\mathbf{x}^{k+1} & =\mathbf{x}^{k}+\gamma^{k}\left(\hat{\mathbf{x}}^{k+1}-\mathbf{x}^{k}\right),
\end{cases}
\end{equation}
where at each iteration, the first stage is generating the descent
direction $\hat{\mathbf{x}}^{k+1}-\mathbf{x}^{k}$, and the second
stage is updating the variable along the solved descent direction
with a step-size $\gamma^{k}$ satisfying 
\begin{equation}
\lim_{k\rightarrow\infty}\gamma^{k}=0\quad\text{and}\quad\sum_{k=0}^{\infty}\gamma^{k}=\infty.
\end{equation}

The generated sequence $\left\{ \mathbf{x}^{k}\right\} $ is proven
to converge to a generalized stationary point of the original problem
(\ref{eq: SCA original problem}) under the following mild assumptions
\cite{facchinei2020ghost}:

\begin{assumption} \label{asm: convergence conditions for general SCA}Let
$O_{\beta}$ and $O_{\mathcal{K}}$ be open neighborhoods of $\left\{ \mathbf{x}\vert\Vert\mathbf{x}-\mathbf{x}^{k}\Vert_{\infty}\le\beta\right\} $
and $\mathcal{K}$ and such that:

\textbf{On original problem} (\ref{eq: SCA original problem}): 

A1) $\mathcal{K}$ is an nonempty, closed, and convex set.

A2) $f\left(\mathbf{x}\right)$ and $g_{i}\left(\mathbf{x}\right)$
are continuously differentiable with locally Lipschitz gradients on
an open set containing $\mathcal{\mathcal{K}}$.

\textbf{On surrogate function} $\tilde{f}$: 

B1) $\tilde{f}\left(\mathbf{x};\mathbf{y}\right)$ is a strongly convex
function on $O_{\beta}$ for every $\mathbf{y}\in\mathcal{K}$ with
modulus of strong convexity $c>0$ independent of $\mathbf{y}$;

B2) $\tilde{f}\left(\mathbf{x};\mathbf{y}\right)$ is continuous on
$O_{\beta}\times O_{\mathcal{K}}$;

B3) $\triangledown_{1}\tilde{f}\left(\mathbf{x};\mathbf{y}\right)$
is continuous on $O_{\beta}\times O_{\mathcal{K}}$;

B4) $\triangledown_{1}\tilde{f}\left(\mathbf{y};\mathbf{y}\right)=\triangledown f\left(\mathbf{y}\right)$
for every $\mathbf{y}\in\mathcal{K}$;

\textbf{On surrogate constraint} $\tilde{g_{i}}$: 

C1) $\tilde{g}_{i}\left(\mathbf{x};\mathbf{y}\right)$ is a convex
function on $O_{\beta}$ for every $\mathbf{y}\in\mathcal{K}$;

C2) $\tilde{g}_{i}\left(\mathbf{x};\mathbf{y}\right)$ is continuous
on $\mathbb{R}^{N}\times O_{\mathcal{K}}$;

C3) $\tilde{g}_{i}\left(\mathbf{x};\mathbf{y}\right)=g_{i}\left(\mathbf{y}\right)$
for every $\mathbf{y}\in\mathcal{K}$;

C4) $\triangledown_{1}\tilde{g}_{i}\left(\mathbf{x};\mathbf{y}\right)$
is continuous on $O_{\beta}\times O_{\mathcal{K}}$;

C5) $\triangledown_{1}\tilde{g}_{i}\left(\mathbf{y};\mathbf{y}\right)=\triangledown f\left(\mathbf{y}\right)$
for every $\mathbf{y}\in\mathcal{K}$;

where $\triangledown_{1}\tilde{f}\left(\mathbf{u};\mathbf{y}\right)$
and $\triangledown_{1}\tilde{g}_{i}\left(\mathbf{u};\mathbf{y}\right)$
denote the partial gradient of $\tilde{f}\left(\mathbf{u};\mathbf{y}\right)$
and $\tilde{g}_{i}\left(\mathbf{u};\mathbf{y}\right)$ evaluated at
$\mathbf{u}$.

\end{assumption}

We can simplify the surrogate problem (\ref{eq: SCA surrogate problem})
accordingly when the following assumptions are additionally satisfied:
\begin{enumerate}
\item if $\mathcal{K}$ is bounded, then the constraint $\Vert\mathbf{x}-\mathbf{x}^{k}\Vert_{\infty}\le\beta$
can be ignored;
\item if $\triangledown f\left(\mathbf{x}\right)$ is Lipschitz continuous
on $\mathcal{K}$ and $\tilde{g}_{i}\left(\mathbf{x};\mathbf{x}^{k}\right)\ge g_{i}\left(\mathbf{x}\right)$
is satisfied for every $\mathbf{x}\in\mathcal{K}$, then the constraint
$\Vert\mathbf{x}-\mathbf{x}^{k}\Vert_{\infty}\le\beta$ can be ignored
and $\eta\left(\mathbf{x}^{k}\right)\equiv0$ \cite{Scutari2016parallel};
\item \textcolor{black}{if $\triangledown f\left(\mathbf{x}\right)$ is
Lipschitz continuous on $\mathcal{K}$ and $\tilde{g}_{i}\left(\mathbf{x};\mathbf{x}^{k}\right)=g_{i}\left(\mathbf{x}\right)$
is satisfied for every $\mathbf{x}\in\mathcal{K}$, then the algorithm
reduces to the vanilla SCA algorithm. The constraint $\Vert\mathbf{x}-\mathbf{x}^{k}\Vert_{\infty}\le\beta$
can be ignored and $\eta\left(\mathbf{x}^{k}\right)\equiv0$ \cite{scutari2013decomposition};}
\item if $\mathcal{K}$ is bounded, $\tilde{f}\left(\mathbf{x};\mathbf{x}^{k}\right)\ge f\left(\mathbf{x}\right)$
and $\tilde{g}_{i}\left(\mathbf{x};\mathbf{x}^{k}\right)=g_{i}\left(\mathbf{x}\right)$
are satisfied for every $\mathbf{x}\in\mathcal{K}$, then the algorithm
reduces to the classical majorization-minimization (MM) method with
convex majorization functions. The constraint $\Vert\mathbf{x}-\mathbf{x}^{k}\Vert_{\infty}\le\beta$
can be ignored, $\eta\left(\mathbf{x}^{k}\right)\equiv0$, and $\gamma^{k}$
can be simply fixed to $1$ \cite{sun2016majorization,razaviyayn2013unified}.
\end{enumerate}

\section{Solving the MVSK Portfolio Problem via SCA \label{sec: solving MVSK}}

In this section, we discuss how to solve the problem (\ref{eq: mvsk portfolio})
via the SCA algorithm. We first investigate the Difference of Convex
(DC) programming approach for solving the problem (\ref{eq: mvsk portfolio})
\cite{dinh2011efficient}, which is actually a special case of the
MM algorithm. Inspired by this, we herein propose another MM based
algorithm by constructing a sequence of tighter upper bound functions.
Thus fewer iterations can be expected. However, we further recognize
that the MM algorithm might still be too conservative as it requires
constructing a global upper for the objective function. Therefore,
we further propose a general SCA based algorithm for solving the problem
(\ref{eq: mvsk portfolio}), where a strongly convex approximating
function is constructed for the objective function.

\subsection{\textcolor{black}{Preliminary Approach: DC Algorithm}}

A DC approach method was proposed in \cite{dinh2011efficient} to
solve problem (\ref{eq: mvsk portfolio}) by recognizing that $\triangledown^{2}f\left(\mathbf{w}\right)$
has a bounded spectral radius under the bounded feasible set $\mathcal{W}$. 
\begin{lem}
\cite{dinh2011efficient} \label{lem: bound for hessian matrix DC alg.}Given
$\mathbf{w}\ge\mathbf{0}$, $\mathbf{1}^{T}\mathbf{w}=1$, we have
\begin{equation}
\begin{aligned}\rho\left(\triangledown^{2}f\left(\mathbf{w}\right)\right) & \le2\lambda_{2}\Vert\boldsymbol{\Sigma}\Vert_{\infty}+6\lambda_{3}\max_{1\le i\le N}\sum_{j,k=1}^{N}\vert\Phi_{ij}^{(k)}\vert\\
 & \quad+12\lambda_{4}\max_{1\le i\le N}\sum_{j,k,l=1}^{N}\vert\Psi_{ij}^{(k,l)}\vert,
\end{aligned}
\end{equation}
where $\rho\left(\mathbf{X}\right)$ is the spectral radius of $\mathbf{X}$.
\end{lem}
The bound for $\rho\left(\triangledown^{2}f\left(\mathbf{w}\right)\right)$
provided in Lemma \ref{lem: bound for hessian matrix DC alg.} can
be easily extended under the constraints in (\ref{eq: constrains for w})
(where instead of no-shorting $\mathbf{w}\ge\mathbf{0}$ we allow
some leverage of $L$ with $\Vert\mathbf{w}\Vert{}_{1}\le L$) to
\[
\begin{aligned}\rho\left(\triangledown^{2}f\left(\mathbf{w}\right)\right) & \le2\lambda_{2}\Vert\boldsymbol{\Sigma}\Vert_{\infty}+6\lambda_{3}L\max_{1\le i\le N}\sum_{j,k=1}^{N}\vert\Phi_{ij}^{(k)}\vert\\
 & \quad+12\lambda_{4}L^{2}\max_{1\le i\le N}\sum_{j,k,l=1}^{N}\vert\Psi_{ij}^{(k,l)}\vert,
\end{aligned}
\]
Then we can represent $f\left(\mathbf{w}\right)$ as 
\begin{equation}
f\left(\mathbf{w}\right)=\frac{\tau_{\text{DC}}}{2}\mathbf{w}^{T}\mathbf{w}-\left(\frac{\tau_{\text{DC}}}{2}\mathbf{w}^{T}\mathbf{w}-f\left(\mathbf{w}\right)\right),
\end{equation}
where both $\frac{\tau_{\text{DC}}}{2}\mathbf{w}^{T}\mathbf{w}$ and
$\frac{\tau_{\text{DC}}}{2}\mathbf{w}^{T}\mathbf{w}-f\left(\mathbf{w}\right)$
are convex functions in $\mathbf{w}$ if $\tau_{\text{DC}}\ge\rho\left(\triangledown^{2}f\left(\mathbf{w}\right)\right)$.
Then the classical concave-convex procedure (CCCP) can be employed
here by iteratively linearizing the second (concave) term, i.e., 
\begin{equation}
\begin{aligned}\underset{\mathbf{w}}{\mathsf{minimize}}\,\,\, & \,\,\,\,\frac{\tau_{\text{DC}}}{2}\mathbf{w}^{T}\mathbf{w}-\mathbf{w}^{T}\left(\tau_{\text{DC}}\mathbf{w}^{k}-\triangledown f\left(\mathbf{w}^{k}\right)\right)\\
\mathsf{subject\,\,to} & \quad\mathbf{w}\in\mathcal{W},
\end{aligned}
\label{eq: DC sub problem}
\end{equation}
where $\triangledown f\left(\mathbf{w}^{k}\right)=-\lambda_{1}\triangledown\phi_{1}\left(\mathbf{w}^{k}\right)+\lambda_{2}\triangledown\phi_{2}\left(\mathbf{w}^{k}\right)-\lambda_{3}\triangledown\phi_{3}\left(\mathbf{w}^{k}\right)+\lambda_{4}\triangledown\phi_{4}\left(\mathbf{w}^{k}\right)$.
It is already a convex problem and can be easily solved. Furthermore,
we can rewrite it as a convex quadratic programing (QP) problem by
introducing a variable $\mathbf{u}\in\mathbb{R}^{N}$:
\begin{equation}
\begin{aligned}\underset{\mathbf{w},\mathbf{u}}{\mathsf{minimize}}\,\,\, & \,\,\,\,\frac{\tau_{\text{DC}}}{2}\mathbf{w}^{T}\mathbf{w}-\mathbf{w}^{T}\left(\tau_{\text{DC}}\mathbf{w}^{k}-\triangledown f\left(\mathbf{w}^{k}\right)\right)\\
\mathsf{subject\,\,to} & \quad\mathbf{1}^{T}\mathbf{w}=1,-\mathbf{u}\le\mathbf{w}\le\mathbf{u},\mathbf{1}^{T}\mathbf{u}\le L,
\end{aligned}
\end{equation}
which can be very efficiently solved with a QP solver. In the rest
of the paper, we will always use this trick to transform the $\ell_{1}$-norm
constraint to linear inequality constraints. The complete DC algorithm
for solving the problem (\ref{eq: mvsk portfolio}) is given in Algorithm
\ref{alg: DC for mvsk portfolio}.
\begin{algorithm}
\caption{DC method for problem (\ref{eq: mvsk portfolio}). \label{alg: DC for mvsk portfolio}}
\begin{algorithmic}[1]

\STATE Initialize $\mathbf{w}^{0}\in\mathcal{W}$ and compute $\tau_{\text{DC}}\ge\rho\left(\triangledown^{2}f\left(\mathbf{w}\right)\right)$
as in Lemma \ref{lem: bound for hessian matrix DC alg.}.

\FOR{$k=0,1,2,\ldots$}

\STATE Calculate $\triangledown f\left(\mathbf{w}^{k}\right)$.

\STATE Solve the problem (\ref{eq: DC sub problem}) to obtain $\hat{\mathbf{w}}^{k+1}$.

\STATE $\mathbf{w}^{k+1}=\hat{\mathbf{w}}^{k+1}$.

\STATE Terminate loop if converges.

\ENDFOR

\end{algorithmic}
\end{algorithm}

\subsection{\textcolor{black}{Preliminary Approach: MM Algorithm} \label{subsec: MM for mvsk}}

The DC algorithm is a special case of the more general MM algorithm,
which works by solving a sequence of global upper bound problems of
the original problem \cite{hunter2004tutorial,sun2016majorization}.
Inspired by the DC approach discussed in the above section, we propose
a tighter upper bound function for $f\left(\mathbf{w}\right)$. Note
that the objective in the surrogate problem (\ref{eq: DC sub problem})
can be rewritten as
\begin{equation}
\begin{aligned} & \frac{\tau_{\text{DC}}}{2}\mathbf{w}^{T}\mathbf{w}-\tau_{\text{DC}}\left(\mathbf{w}^{k}\right)^{T}\mathbf{w}+\triangledown f\left(\mathbf{w}^{k}\right)^{T}\mathbf{w}+\text{const.}\\
 & =f\left(\mathbf{w}^{k}\right)+\triangledown f\left(\mathbf{w}^{k}\right)^{T}\left(\mathbf{w}-\mathbf{w}^{k}\right)+\frac{\tau_{\text{DC}}}{2}\Vert\mathbf{w}-\mathbf{w}^{k}\Vert_{2}^{2},
\end{aligned}
\end{equation}
It is actually a global upper bound function of $f\left(\mathbf{w}\right)$
\cite{sun2016majorization} at $\mathbf{w}^{k}$. However, denoting
$f\left(\mathbf{w}\right)=f_{\text{cvx}}\left(\mathbf{w}\right)+f_{\text{ncvx}}\left(\mathbf{w}\right)$
with $f_{\text{cvx}}\left(\mathbf{w}\right)=-\lambda_{1}\phi_{1}\left(\mathbf{w}\right)+\lambda_{2}\phi\left(\mathbf{w}\right)$
and $f_{\text{ncvx}}\left(\mathbf{w}\right)=-\lambda_{3}\phi_{3}\left(\mathbf{w}\right)+\lambda_{4}\phi_{4}\left(\mathbf{w}\right)$,
we find $f_{\text{cvx}}\left(\mathbf{w}\right)$ is already a convex
function. Then we can merely construct the an upper bound function
for $f_{\text{ncvx}}\left(\mathbf{w}\right)$. Inspired by Lemma \ref{lem: bound for hessian matrix DC alg.},
we propose a smaller bound for $\rho\left(\triangledown^{2}f_{\text{ncvx}}\left(\mathbf{w}\right)\right)$
as follows.
\begin{lem}
\label{lem: bound for hessian matrix MM alg.}Under the constraints
in (\ref{eq: constrains for w}), we have
\begin{equation}
\begin{aligned}\rho\left(\triangledown^{2}f_{\text{ncvx}}\left(\mathbf{w}\right)\right) & \le6\lambda_{3}L\max_{1\le i\le N}\sum_{j=1}^{N}\max_{1\le k\le N}\vert\Phi_{ij}^{(k)}\vert\\
 & \quad+12\lambda_{4}L^{2}\max_{1\le i\le N}\sum_{j=1}^{N}\max_{1\le k,l\le N}\vert\Psi_{ij}^{(k,l)}\vert.
\end{aligned}
\end{equation}
\end{lem}
\begin{IEEEproof}
See Appendix \ref{apx: proof MM hsn bound}.
\end{IEEEproof}
Then we can construct, compared with the upper bound function actually
used in DC method, a much tighter upper bound function $\check{f}_{\text{ncvx}}\left(\mathbf{w}\right)$
for $f_{\text{ncvx}}\left(\mathbf{w}\right)$ at $\mathbf{w}^{k}$
as \cite{sun2016majorization}:
\begin{equation}
\begin{aligned}\check{f}_{\text{ncvx}}\left(\mathbf{w},\mathbf{w}^{k}\right) & =f_{\text{ncvx}}\left(\mathbf{w}^{k}\right)+\triangledown f_{\text{ncvx}}\left(\mathbf{w}^{k}\right)^{T}\left(\mathbf{w}-\mathbf{w}^{k}\right)\\
 & \quad+\frac{\tau_{\text{MM}}}{2}\Vert\mathbf{w}-\mathbf{w}^{k}\Vert_{2}^{2},
\end{aligned}
\end{equation}
where $\triangledown f_{\text{ncvx}}\left(\mathbf{w}^{k}\right)=-\lambda_{3}\triangledown\phi_{3}\left(\mathbf{w}^{k}\right)+\lambda_{4}\triangledown\phi_{4}\left(\mathbf{w}^{k}\right)$
and $\tau_{\text{MM}}\ge\rho\left(\triangledown^{2}f_{\text{ncvx}}\left(\mathbf{w}\right)\right)$
can be calculated via Lemma \ref{lem: bound for hessian matrix MM alg.}.
Then a tighter global upper bound function can be constructed for
$f\left(\mathbf{w}\right)$ as $\check{f}\left(\mathbf{w},\mathbf{w}^{k}\right)=f_{\text{cvx}}\left(\mathbf{w}\right)+\check{f}_{\text{ncvx}}\left(\mathbf{w},\mathbf{w}^{k}\right)$.
At each iteration of the MM algorithm, we need solve the following
surrogate problem:
\begin{equation}
\begin{aligned}\underset{\mathbf{w}}{\mathsf{minimize}}\,\,\, & \quad\mathbf{w}^{T}\check{\mathbf{Q}}^{k}\mathbf{w}+\mathbf{w}^{T}\check{\mathbf{q}}^{k}\\
\mathsf{subject\,\,to} & \quad\mathbf{w}\in\mathcal{W},
\end{aligned}
\label{eq: MM sub problem}
\end{equation}
where $\check{\mathbf{Q}}^{k}=\lambda_{2}\boldsymbol{\Sigma}+\frac{\tau_{\text{MM}}}{2}\mathbf{I}$
and $\check{\mathbf{q}}^{k}=-\lambda_{1}\boldsymbol{\mu}+\triangledown f_{\text{ncvx}}\left(\mathbf{w}^{k}\right)-\tau_{\text{MM}}\mathbf{w}^{k}$.
It is a strongly convex QP problem and can be very efficiently solved
by a QP solver. The complete MM algorithm for solving the problem
(\ref{eq: mvsk portfolio}) is given in Algorithm \ref{alg: MM for mvsk portfolio}.
Compared with the original DC algorithm, the MM algorithm does not
introduce any additional computation, while we can expect faster convergence.
\begin{algorithm}
\caption{MM method for problem (\ref{eq: mvsk portfolio}). \label{alg: MM for mvsk portfolio}}
\begin{algorithmic}[1]

\STATE Initialize $\mathbf{w}^{0}\in\mathcal{W}$ and compute $\tau_{\text{MM}}\ge\rho\left(\triangledown^{2}f_{\text{ncvx}}\left(\mathbf{w}\right)\right)$
as in Lemma \ref{lem: bound for hessian matrix MM alg.}.

\FOR{$k=0,1,2,\ldots$}

\STATE Calculate $\triangledown f_{\text{ncvx}}\left(\mathbf{w}^{k}\right)$.

\STATE Solve the problem (\ref{eq: MM sub problem}) to obtain $\hat{\mathbf{w}}^{k+1}$.

\STATE $\mathbf{w}^{k+1}=\hat{\mathbf{w}}^{k+1}$.

\STATE Terminate loop if converges.

\ENDFOR

\end{algorithmic}
\end{algorithm}

\subsection{\textcolor{black}{Q-MVSK Algorithm} \label{subsec: SCA for mvsk}}

\begin{figure}
\begin{centering}
\includegraphics[scale=0.5]{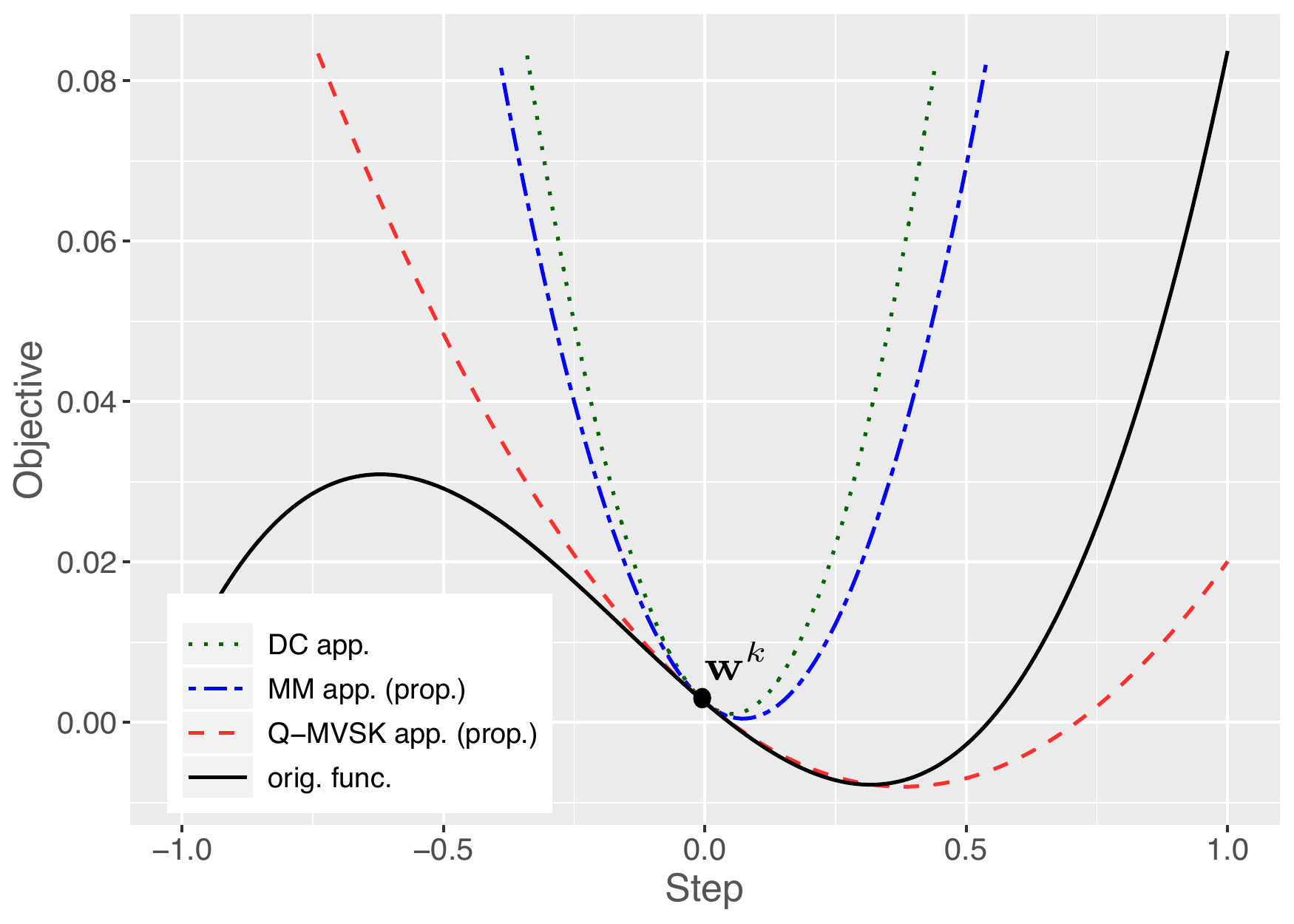}
\par\end{centering}
\caption{Illustration of approximating functions. \label{fig: Illustration-of-approximation}}

\end{figure}

The MM-type methods require constructing a global upper bound approximation,
which is sometimes criticized to be too conservative to capture the
global landscape for the objective function \cite{nedic2018multi}.
Therefore, in this section, we propose the Q-MVSK algorithm to solve
the problem (\ref{eq: mvsk portfolio}) via a strongly convex approximation
(need not be a global upper bound) for the objective. More specifically,
we still leave the convex part $f_{\text{cvx}}\left(\mathbf{w}\right)$
untouched but construct a second-order approximation for $f_{\text{ncvx}}\left(\mathbf{w}\right)$
as
\begin{equation}
\begin{aligned} & \tilde{f}_{\text{ncvx}}\left(\mathbf{w},\mathbf{w}^{k}\right)\\
 & =f_{\text{ncvx}}\left(\mathbf{w}^{k}\right)+\triangledown f_{\text{ncvx}}\left(\mathbf{w}^{k}\right)^{T}\left(\mathbf{w}-\mathbf{w}^{k}\right)\\
 & \quad+\frac{1}{2}\left(\mathbf{w}-\mathbf{w}^{k}\right)^{T}\mathbf{H}_{\text{ncvx}}^{k}\left(\mathbf{w}-\mathbf{w}^{k}\right)+\frac{\tau_{\mathbf{w}}}{2}\Vert\mathbf{w}-\mathbf{w}^{k}\Vert_{2}^{2},
\end{aligned}
\end{equation}
where $\mathbf{H}_{\text{ncvx}}^{k}$ is an approximation of $\triangledown^{2}f_{\text{ncvx}}\left(\mathbf{w}^{k}\right)$
with $\triangledown^{2}f_{\text{ncvx}}\left(\mathbf{w}^{k}\right)=-\lambda_{3}\triangledown^{2}\phi_{3}\left(\mathbf{w}^{k}\right)+\lambda_{4}\triangledown^{2}\phi_{4}\left(\mathbf{w}^{k}\right)$
from Lemma \ref{lem: gradient and hessian of moments}, and $\tau_{\mathbf{w}}\ge0$
is to preserve the strong convexity of $\tilde{f}_{\text{ncvx}}\left(\mathbf{w},\mathbf{w}^{k}\right)$.
Note that $\tau_{\mathbf{w}}$ can be set $0$ to when $\lambda_{2}>0$.
$\mathbf{H}_{\text{ncvx}}^{k}$ is a positive semidefinite matrix
close to $\triangledown^{2}f_{\text{ncvx}}\left(\mathbf{w}^{k}\right)$
obtained as follows.
\begin{lem}
\label{lem: nearest PSD} \cite{higham1988computing} The nearest
symmetric positive semidefinite matrix in the Frobenius norm to a
real symmetric real matrix $\mathbf{X}$ is $\mathbf{U}\mathsf{Diag}\left(\mathbf{d}_{+}\right)\mathbf{U}^{T}$,
where $\mathbf{U}\mathsf{Diag}\left(\mathbf{d}\right)\mathbf{U}^{T}$
is the eigenvalue decomposition of $\mathbf{X}$.
\end{lem}
Then we have an approximating function for $f\left(\mathbf{w}\right)$
as $\tilde{f}\left(\mathbf{w},\mathbf{w}^{k}\right)=f_{\text{cvx}}\left(\mathbf{w}\right)+\tilde{f}_{\text{ncvx}}\left(\mathbf{w},\mathbf{w}^{k}\right)$.
In Figure \ref{fig: Illustration-of-approximation}, the three approximating
functions are illustrated by being restricted to a line on $\mathcal{W}$.
We can see that $\tilde{f}\left(\mathbf{w},\mathbf{w}^{k}\right)$
can best describe the global behaviour of $f\left(\mathbf{w}\right)$.
 At each iteration of the MM algorithm, we need solve the following
surrogate problem:
\begin{equation}
\begin{aligned}\underset{\mathbf{w}}{\mathsf{minimize}}\,\,\, & \quad\mathbf{w}^{T}\tilde{\mathbf{Q}}^{k}\mathbf{w}+\mathbf{w}^{T}\tilde{\mathbf{q}}^{k}\\
\mathsf{subject\,\,to} & \quad\mathbf{w}\in\mathcal{W},
\end{aligned}
\label{eq: SCA sub problem}
\end{equation}
where $\tilde{\mathbf{Q}}^{k}=\lambda_{2}\boldsymbol{\Sigma}+\frac{1}{2}\mathbf{H}_{\text{ncvx}}^{k}+\frac{\tau_{\mathbf{w}}}{2}\mathbf{I}$
and $\tilde{\mathbf{q}}^{k}=-\lambda_{1}\boldsymbol{\mu}+\triangledown f_{\text{ncvx}}\left(\mathbf{w}^{k}\right)-\mathbf{H}_{\text{ncvx}}^{k}\mathbf{w}^{k}-\tau_{\mathbf{w}}\mathbf{w}^{k}$.
It is a strongly convex QP problem and can be very efficiently solved
by a QP solver. The complete Q-MVSK algorithm for solving the problem
(\ref{eq: mvsk portfolio}) is given in Algorithm \ref{alg: SCA for mvsk portfolio}.
\begin{algorithm}
\caption{Q-MVSK algorithm for problem (\ref{eq: mvsk portfolio}). \label{alg: SCA for mvsk portfolio}}
\begin{algorithmic}[1]

\STATE Initialize $\mathbf{w}^{0}\in\mathcal{W}$ and pick a sequence
$\left\{ \gamma^{k}\right\} $.

\FOR{$k=0,1,2,\ldots$}

\STATE Calculate $\triangledown f_{\text{ncvx}}\left(\mathbf{w}^{k}\right)$,
$\mathbf{H}_{\text{ncvx}}^{k}$.

\STATE Solve the problem (\ref{eq: SCA sub problem}) to obtain $\hat{\mathbf{w}}^{k+1}$.

\STATE $\mathbf{w}^{k+1}=\mathbf{w}^{k}+\gamma^{k}\left(\hat{\mathbf{w}}^{k+1}-\mathbf{w}^{k}\right)$.

\STATE Terminate loop if converges.

\ENDFOR

\end{algorithmic}
\end{algorithm}

\section{Solving The MVSK Tilting Portfolio Problem via SCA \label{sec: solving MVSK tilting}}

In this section, we discuss how to solve the MVSK tilting problem
(\ref{eq: mvsk tilting problem}), which we rewrite as
\begin{equation}
\begin{aligned}\underset{\mathbf{w},\delta}{\mathsf{minimize}}\,\,\, & \quad-\delta\\
\mathsf{subject\,\,to} & \quad g_{i}\left(\mathbf{w},\delta\right)\le0,\,\,i=1,\dots,5\\
 & \quad\mathbf{w}\in\mathcal{W},\delta\ge0,
\end{aligned}
\label{eq: mvsk tilting smooth}
\end{equation}
where 
\begin{equation}
\begin{aligned}g_{1}\left(\mathbf{w},\delta\right) & =\phi_{1}\left(\mathbf{w}_{0}\right)-\phi_{1}\left(\mathbf{w}\right)+d_{1}\delta,\\
g_{2}\left(\mathbf{w},\delta\right) & =\phi_{2}\left(\mathbf{w}\right)-\phi_{2}\left(\mathbf{w}_{0}\right)+d_{2}\delta,\\
g_{3}\left(\mathbf{w},\delta\right) & =\phi_{3}\left(\mathbf{w}_{0}\right)-\phi_{3}\left(\mathbf{w}\right)+d_{3}\delta,\\
g_{4}\left(\mathbf{w},\delta\right) & =\phi_{4}\left(\mathbf{w}\right)-\phi_{4}\left(\mathbf{w}_{0}\right)+d_{4}\delta,\\
g_{5}\left(\mathbf{w},\delta\right) & =\left(\mathbf{w}-\mathbf{w}_{\text{ref}}\right)^{T}\boldsymbol{\Sigma}\left(\mathbf{w}-\mathbf{w}_{\text{ref}}\right)-\kappa^{2}.
\end{aligned}
\end{equation}
Note that $g_{i}\left(\mathbf{w},\delta\right),i=1,2,5$ are all convex
functions, while $g_{i}\left(\mathbf{w},\delta\right),i=3,4$ are
both nonconvex functions. We will next explore several options to
deal with problem (\ref{eq: mvsk tilting smooth}), which contains
nonconvex constraints.

The classical way for solving such constrained problem is the interior-point
(a.k.a. barrier) method (IPM), which adds the indicator functions
for the inequality constraints to the objective and approximates them
with logarithmic barrier functions \cite{boyd2004convex}. The IPM
method can be employed to the problem (\ref{eq: mvsk tilting problem})
and transform it to 
\begin{equation}
\begin{aligned}\underset{\mathbf{w},\delta}{\mathsf{minimize}}\,\,\, & \quad-t\delta-\sum_{i=1}^{5}\log\left(-g_{i}\left(\mathbf{w},\delta\right)\right)\\
\mathsf{subject\,\,to} & \quad\mathbf{w}\in\mathcal{W},\delta\ge0,
\end{aligned}
\label{eq: mvsk tilting IPM}
\end{equation}
where $t>0$ is a parameter that sets the accuracy of the barrier
approximation. Then we could solve the problem (\ref{eq: mvsk tilting IPM})
via a general gradient descend method or SCA algorithm. However,\textcolor{black}{{}
due to the implicit constraint $g_{i}\left(\mathbf{w},\delta\right)\le0$,
a} line search is compulsory at each iteration to guarantee a feasible
update of $\left(\mathbf{w},\delta\right)$. As we have discussed
before, the computational complexity of a single evaluation of $g_{4}\left(\mathbf{w},\delta\right)$
is $\mathcal{O}\left(N^{4}\right)$. Then the line search is too computationally
expensive to be practical in this problem.

Another way to solve problem (\ref{eq: mvsk tilting smooth}) could
be by constructing a global upper bound approximation for all the
nonconvex constraints and solve a sequence of inner convex approximating
problems. Using the upper bound construction procedure in Section
\ref{subsec: MM for mvsk}, we can easily construct an inner convex
approximating problem for problem (\ref{eq: mvsk tilting smooth})
at $\mathbf{w}^{k}$ as:
\begin{equation}
\begin{aligned}\underset{\mathbf{w},\delta}{\mathsf{minimize}}\,\,\, & \quad-\delta+\frac{\tau_{\delta}}{2}(\delta-\delta^{k})^{2}+\frac{\tau_{\mathbf{w}}}{2}\Vert\mathbf{w}-\mathbf{w}^{k}\Vert_{2}^{2}\\
\mathsf{subject\,\,to} & \quad g_{i}\left(\mathbf{w},\delta\right)\le0,\,\,i=1,2,5,\\
 & \quad\check{g}_{j}\left(\mathbf{w},\delta;\mathbf{w}^{k},\delta^{k}\right)\le0,\,\,j=3,4,\\
 & \quad\mathbf{w}\in\mathcal{W},\delta\ge0,
\end{aligned}
\label{eq: mvsk tilting inner appro}
\end{equation}
where $\check{g}_{j}\left(\mathbf{w},\delta;\mathbf{w}^{k},\delta^{k}\right)$
is the global upper bound of $g_{j}\left(\mathbf{w},\delta\right)$
at $\left(\mathbf{w}^{k},\delta^{k}\right)$, which can be constructed
as in Section \ref{subsec: MM for mvsk}. The problem (\ref{eq: mvsk tilting inner appro})
is a convex quadratically constrained quadratic programing (QCQP)
problem and can be solved via several solvers. However, we can know
from Figure \ref{fig: Illustration-of-approximation} and the numerical
experiments in Section \ref{subsec: num res for mvsk solving} that
such upper bound is very loose and the convergence is slow. 

\textcolor{black}{Instead, we proposed constructing convex approximations
(although not upper bounds) for the nonconvex constraints in the following.}

\subsection{\textcolor{black}{Preliminary Approach: L-MVSKT Algorithm} \label{subsec: L-MVSKT}}

The most classical choice, as mentioned in \cite{facchinei2020ghost},
is approximating the objective function by a quadratic function while
linearizing all constraints. Therefore, we herein propose the L-MVSKT
algorithm by linearizing all the non-linear constraints in problem
(\ref{eq: mvsk tilting problem}), i.e., the surrogate problem is
\begin{equation}
\begin{aligned}\underset{\mathbf{w},\delta}{\mathsf{minimize}}\,\,\, & \quad-\delta+\frac{\tau_{\delta}}{2}(\delta-\delta^{k})^{2}+\frac{\tau_{\mathbf{w}}}{2}\Vert\mathbf{w}-\mathbf{w}^{k}\Vert_{2}^{2}\\
\mathsf{subject\,\,to} & \quad g_{1}\left(\mathbf{w},\delta\right)\le0\\
 & \quad\bar{g}_{j}(\mathbf{w},\delta;\mathbf{w}^{k},\delta^{k})\le\eta(\mathbf{w}^{k},\delta^{k}),j=2,3,4,5\\
 & \quad\mathbf{w}\in\mathcal{W},\delta\ge0,
\end{aligned}
\label{eq: mvsk tilting linecon surrogate problem}
\end{equation}
where $\bar{g}_{j}(\mathbf{w},\delta;\mathbf{w}^{k},\delta^{k})$
is the linear approximation of $g_{j}\left(\mathbf{w},\delta\right)$
at $\left(\mathbf{w}^{k},\delta^{k}\right)$ with
\begin{equation}
\begin{aligned} & \bar{g}_{j}(\mathbf{w},\delta;\mathbf{w}^{k},\delta^{k})\\
 & =g_{1}\left(\mathbf{w}^{k},\delta^{k}\right)+\triangledown_{\mathbf{w}}g_{j}\left(\mathbf{w}^{k},\delta^{k}\right)^{T}\left(\mathbf{w}-\mathbf{w}^{k}\right)\\
 & \quad+\triangledown_{\delta}g_{j}\left(\mathbf{w}^{k},\delta^{k}\right)^{T}(\delta-\delta^{k}),j=2,3,4,
\end{aligned}
\end{equation}
Besides, $\eta\left(\mathbf{w}^{k},\delta^{k}\right)$ here can be
computed as
\begin{equation}
\begin{aligned} & \eta\left(\mathbf{w}^{k},\delta^{k}\right)\\
 & \triangleq\left(1-\theta\right)\max_{j=2,3,4,5}\left\{ g_{j}\left(\mathbf{w}^{k},\delta^{k}\right)_{+}\right\} \\
 & \quad+\theta\min_{\mathbf{w},\delta}\left\{ \max_{j=2,3,4,5}\big\{\bar{g}_{j}\left(\mathbf{w},\delta;\mathbf{w}^{k},\delta^{k}\right)_{+}\big\}\big|\left(\mathbf{w},\delta\right)\in\bar{\mathcal{W}}\right\} ,
\end{aligned}
\label{eq: eta for mvsk tilting linecon}
\end{equation}
where $\bar{\mathcal{W}}$ is a convex set defined as
\begin{equation}
\begin{aligned}\bar{\mathcal{W}} & =\Big\{\left(\mathbf{w},\delta\right)\big|\mathbf{w}\in\mathcal{W},g_{1}\left(\mathbf{w},\delta\right)\le0,\delta\ge0\Big\}\end{aligned}
.
\end{equation}
\textcolor{black}{The second term in equation (\ref{eq: eta for mvsk tilting linecon})
is obtained as $t$ from solving the following problem:
\begin{equation}
\begin{aligned}\underset{\mathbf{w},\delta,t}{\mathsf{minimize}}\,\,\, & \quad t\\
\mathsf{subject\,\,to} & \quad\bar{g}_{j}\left(\mathbf{w},\delta;\mathbf{w}^{k},\delta^{k}\right)\le t,\,\,j=2,3,4,5,\\
 & \quad\left(\mathbf{w},\delta\right)\in\bar{\mathcal{W}},t\ge0.
\end{aligned}
\label{eq: eta sub problem for mvsk linecon}
\end{equation}
}Problem (\ref{eq: mvsk tilting linecon surrogate problem}) is a
convex QP problem and problem (\ref{eq: eta sub problem for mvsk linecon})
is a linear programing (LP) problem. Both of them can be very efficiently
solved by a QP solver and an LP solver, respectively. The complete
L-MVSKT algorithm is given in the Algorithm \ref{alg: SCA for mvsk tilting linecon}.
\begin{algorithm}
\caption{L-MVSKT algorithm for problem (\ref{eq: mvsk tilting problem}). \label{alg: SCA for mvsk tilting linecon}}
\begin{algorithmic}[1]

\STATE Initialize $\mathbf{w}^{0}\in\mathcal{W}$ and pick $\tau_{\delta}$,
$\tau_{\mathbf{w}}$ and a sequence $\left\{ \gamma^{k}\right\} $.

\FOR{$k=0,1,2,\ldots$}

\STATE Calculate $\triangledown\phi_{3}\left(\mathbf{w}^{k}\right)$,
$\triangledown\phi_{4}\left(\mathbf{w}^{k}\right)$.

\STATE Solve problem (\ref{eq: eta sub problem for mvsk linecon})
and compute $\eta\left(\mathbf{w}^{k},\delta^{k}\right)$ as in (\ref{eq: eta for mvsk tilting linecon}).

\STATE Solve problem (\ref{eq: mvsk tilting linecon surrogate problem})
to obtain $\hat{\mathbf{w}}^{k+1}$.

\STATE $\mathbf{w}^{k+1}=\mathbf{w}^{k}+\gamma^{k}\left(\hat{\mathbf{w}}^{k+1}-\mathbf{w}^{k}\right)$.

\STATE Terminate loop if converges.

\ENDFOR

\end{algorithmic}
\end{algorithm}

\subsection{\textcolor{black}{Q-MVSKT Algorithm} \label{subsec: Q-MVSKT}}

In the above section, we have proposed the L-MVSKT algorithm for solving
the MVSK tilting problem (\ref{eq: mvsk tilting problem}). However,
it requires us to linearize the tractable convex quadratic constraints
and the simple linearization is rarely regarded as a proper approximation
for nonconvex constraints. In Section \ref{subsec: SCA for mvsk},
we have proposed a quadratic approximation for the third and fourth
central moments. It shows great advantages from the numerical experiments
presented in Section \ref{subsec: num res for mvsk solving}. Therefore,
similar to Section \ref{subsec: SCA for mvsk}, we can construct a
quadratic approximation for the nonconvex constraints in problem (\ref{eq: mvsk tilting problem})
while not approximating the already convex constraints, i.e., 
\begin{equation}
\begin{aligned}\underset{\mathbf{w},\delta}{\mathsf{minimize}}\,\,\, & \quad-\delta+\frac{\tau_{\delta}}{2}(\delta-\delta^{k})^{2}+\frac{\tau_{\mathbf{w}}}{2}\Vert\mathbf{w}-\mathbf{w}^{k}\Vert_{2}^{2}\\
\mathsf{subject\,\,to} & \quad g_{i}\left(\mathbf{w},\delta\right)\le0,\,\,i=1,2,5,\\
 & \quad\tilde{g}_{j}\left(\mathbf{w},\delta;\mathbf{w}^{k},\delta^{k}\right)\le\eta\left(\mathbf{w}^{k},\delta^{k}\right),\,\,j=3,4,\\
 & \quad\mathbf{w}\in\mathcal{W},\delta\ge0,
\end{aligned}
\label{eq: mvsk tilting smooth equ surrogate}
\end{equation}
where $\tilde{g}_{j}\left(\mathbf{w},\delta;\mathbf{w}^{k},\delta^{k}\right)$
is the quadratic approximating function of $g_{j}\left(\mathbf{w},\delta\right)$
at $\left(\mathbf{w}^{k},\delta^{k}\right)$:
\begin{equation}
\begin{aligned} & \tilde{g}_{3}(\mathbf{w},\delta;\mathbf{w}^{k},\delta^{k})\\
 & =\phi_{3}\left(\mathbf{w}_{0}\right)-\phi_{3}(\mathbf{w}^{k})+d_{3}\delta-\triangledown\phi_{3}(\mathbf{w}^{k})^{T}(\mathbf{w}-\mathbf{w}^{k})\\
 & \quad+\frac{1}{2}(\mathbf{w}-\mathbf{w}^{k})^{T}\mathbf{H}_{\Phi}^{k}(\mathbf{w}-\mathbf{w}^{k}),\\
 & \tilde{g}_{4}(\mathbf{w},\delta;\mathbf{w}^{k},\delta^{k})\\
 & =\phi_{4}\left(\mathbf{w}^{k}\right)-\phi_{4}\left(\mathbf{w}_{0}\right)+d_{4}\delta+\triangledown\phi_{4}(\mathbf{w}^{k})^{T}(\mathbf{w}-\mathbf{w}^{k})\\
 & \quad+\frac{1}{2}(\mathbf{w}-\mathbf{w}^{k})^{T}\mathbf{H}_{\Psi}^{k}(\mathbf{w}-\mathbf{w}^{k}),
\end{aligned}
\end{equation}
with $\mathbf{H}_{\Phi}^{k}$ and $\mathbf{H}_{\Psi}^{k}$ being the
PSD approximating matrixes for $-\triangledown^{2}\phi_{3}\left(\mathbf{w}^{k}\right)$
and $\triangledown^{2}\phi_{4}\left(\mathbf{w}^{k}\right)$. $\eta\left(\mathbf{w}^{k},\delta^{k}\right)$
can be computed from
\begin{equation}
\begin{aligned} & \eta\left(\mathbf{w}^{k},\delta^{k}\right)\\
 & \triangleq\left(1-\theta\right)\max_{j=3,4}\left\{ g_{j}\left(\mathbf{w}^{k},\delta^{k}\right)_{+}\right\} \\
 & \quad+\theta\min_{\mathbf{w},\delta}\left\{ \max_{j=3,4}\big\{\tilde{g}_{j}\left(\mathbf{w},\delta;\mathbf{w}^{k},\delta^{k}\right)_{+}\big\}\big|\left(\mathbf{w},\delta\right)\in\tilde{\mathcal{W}}\right\} ,
\end{aligned}
\label{eq: eta for mvsk tilting quadcon}
\end{equation}
where $\tilde{\mathcal{W}}$ is a convex set defined as
\begin{equation}
\begin{aligned}\tilde{\mathcal{W}} & =\Big\{\left(\mathbf{w},\delta\right)\big|\mathbf{w}\in\mathcal{W},\delta\ge0,g_{i}\left(\mathbf{w},\delta\right)\le0,i=1,2,5\Big\}\end{aligned}
\end{equation}
\textcolor{black}{The second term in equation (\ref{eq: eta for mvsk tilting quadcon})
is obtained as $t$ from solving the following problem:}
\begin{equation}
\begin{aligned}\underset{\mathbf{w},\delta,t}{\mathsf{minimize}}\,\,\, & \quad t\\
\mathsf{subject\,\,to} & \quad\tilde{g}_{3}\left(\mathbf{w},\delta;\mathbf{w}^{k},\delta^{k}\right)\le t,\\
 & \quad\tilde{g}_{4}\left(\mathbf{w},\delta;\mathbf{w}^{k},\delta^{k}\right)\le t,\\
 & \quad\left(\mathbf{w},\delta\right)\in\tilde{\mathcal{W}},t\ge0.
\end{aligned}
\label{eq: eta sub problem for mvsk quadcons}
\end{equation}
Problems (\ref{eq: mvsk tilting smooth equ surrogate}) and (\ref{eq: eta sub problem for mvsk quadcons})
are both convex QCQP problems and can be efficiently solved by the
corresponding solvers. We call it the Q-MVSKT algorithm and give the
complete description in Algorithm \ref{alg: SCA for mvsk tilting quadcon}.
\begin{algorithm}
\caption{Q-MVSKT algorithm for problem (\ref{eq: mvsk tilting problem}). \label{alg: SCA for mvsk tilting quadcon}}
\begin{algorithmic}[1]

\STATE Initialize $\mathbf{w}^{0}\in\mathcal{W}$ and pick $\tau_{\delta}$,
$\tau_{\mathbf{w}}$ and a sequence $\left\{ \gamma^{k}\right\} $.

\FOR{$k=0,1,2,\ldots$}

\STATE Calculate $\triangledown\phi_{3}\left(\mathbf{w}^{k}\right)$,
$\triangledown\phi_{4}\left(\mathbf{w}^{k}\right)$, $\mathbf{H}_{\Phi}^{k}$,
and $\mathbf{H}_{\Psi}^{k}$.

\STATE Solve problem (\ref{eq: eta sub problem for mvsk quadcons})
and compute $\eta\left(\mathbf{w}^{k},\delta^{k}\right)$ as in (\ref{eq: eta for mvsk tilting quadcon}).

\STATE Solve problem (\ref{eq: mvsk tilting smooth equ surrogate})
to obtain $\hat{\mathbf{w}}^{k+1}$.

\STATE $\mathbf{w}^{k+1}=\mathbf{w}^{k}+\gamma^{k}\left(\hat{\mathbf{w}}^{k+1}-\mathbf{w}^{k}\right)$.

\STATE Terminate loop if converges.

\ENDFOR

\end{algorithmic}
\end{algorithm}

\section{Complexity and Convergence Analysis \label{sec: compl and convg anal}}

\subsection{Complexity Analysis}

First of all, it should be noted that the memory complexity for solving
the high-order portfolio optimization problem is $\mathcal{O}\left(N^{4}\right)$
as the kurtosis matrix $\boldsymbol{\Psi}$ is of dimension $N\times N^{3}$.
For example, when $N=200$, storing a complete $\boldsymbol{\Psi}$
takes almost $12\,\mathsf{GB}$ memory size. Thus it is impractical
to solve a very large-scale high-order portfolio optimization problem
due to the memory restriction. All the algorithms investigated or
proposed in this paper are iterative methods. Therefore, we discuss
the computational complexity of constructing the surrogate problems
in each iteration, while the computational complexity of solving them
depends on the specific solvers.

\subsubsection{On Solving The MVSK Portfolio Problem (\ref{eq: mvsk portfolio})}

For Algorithm \ref{alg: DC for mvsk portfolio} and \ref{alg: MM for mvsk portfolio},
the per-iteration computational cost of constructing the surrogate
problems comes mainly from computing the gradients, which is $\mathcal{O}\left(N^{4}\right)$.
For Algorithm \ref{alg: SCA for mvsk portfolio}, it is mainly from
computing the gradient $\triangledown f_{\text{ncvx}}\left(\mathbf{w}^{k}\right)$
and Hessian $\triangledown^{2}f_{\text{ncvx}}\left(\mathbf{w}^{k}\right)$,
which in principle are $\mathcal{O}\left(N^{4}\right)$ and $\mathcal{O}\left(N^{5}\right)$,
respectively. However, we can simplify the computation by first computing
$\triangledown^{2}f_{\text{ncvx}}\left(\mathbf{w}^{k}\right)=-\lambda_{3}\triangledown^{2}\phi_{3}\left(\mathbf{w}^{k}\right)+\lambda_{4}\triangledown^{2}\phi_{4}\left(\mathbf{w}^{k}\right)$
as
\begin{equation}
\triangledown^{2}\phi_{3}\left(\mathbf{w}\right)=6\boldsymbol{\Phi}\left(\mathbf{I}\otimes\mathbf{w}\right)=6\left[\boldsymbol{\Phi}^{(1)}\mathbf{w}\,\,\cdots\,\,\boldsymbol{\Phi}^{(N)}\mathbf{w}\right],\,\,\,\,
\end{equation}
\begin{equation}
\begin{aligned}\triangledown^{2}\phi_{4}\left(\mathbf{w}\right) & =12\boldsymbol{\Psi}\left(\mathbf{I}\otimes\mathbf{w}\otimes\mathbf{w}\right)\\
 & =12\left[\boldsymbol{\Psi}^{(1)}\left(\mathbf{w}\otimes\mathbf{w}\right)\,\,\cdots\,\,\boldsymbol{\Psi}^{(N)}\left(\mathbf{w}\otimes\mathbf{w}\right)\right],
\end{aligned}
\end{equation}
where $\boldsymbol{\Phi}^{(i)}$ is the $i$-th block matrix of dimension
$N\times N$ in $\boldsymbol{\Phi}$ and $\boldsymbol{\Psi}^{(i)}$
is the $i$-th block matrix of dimension $N\times N^{2}$ in $\boldsymbol{\Psi}$.
Then the computational complexity of computing $\triangledown^{2}f_{\text{ncvx}}\left(\mathbf{w}^{k}\right)$
is reduced to $\mathcal{O}\left(N^{4}\right)$. With the usage of
Corollary \ref{cor: grad hsn relations}, $\triangledown f_{\text{ncvx}}\left(\mathbf{w}^{k}\right)$
can be easily computed as
\begin{equation}
\triangledown f_{\text{ncvx}}\left(\mathbf{w}^{k}\right)=-\frac{\lambda_{3}}{2}\triangledown^{2}\phi_{3}\left(\mathbf{w}^{k}\right)\mathbf{w}^{k}+\frac{\lambda_{4}}{3}\triangledown^{2}\phi_{4}\left(\mathbf{w}^{k}\right)\mathbf{w}^{k}.
\end{equation}
Then the overall computational complexity of $\triangledown f_{\text{ncvx}}\left(\mathbf{w}^{k}\right)$
and $\triangledown^{2}f_{\text{ncvx}}\left(\mathbf{w}^{k}\right)$
is still $\mathcal{O}\left(N^{4}\right)$. Therefore, the per-iteration
computational cost of constructing the surrogate problems for Algorithms
\ref{alg: DC for mvsk portfolio}, \ref{alg: MM for mvsk portfolio},
and \ref{alg: SCA for mvsk portfolio} are $\mathcal{O}\left(N^{4}\right)$.

\subsubsection{On Solving The MVSK Tilting Portfolio Problem (\ref{eq: mvsk tilting problem})}

The per-iteration computational cost of constructing the surrogate
problems in Algorithm \ref{alg: SCA for mvsk tilting linecon} comes
mainly from computing the gradients, while that in Algorithm \ref{alg: SCA for mvsk tilting quadcon}
from computing both the gradients and Hessian. Similar to the above
analysis, the latter can be simplified so that both algorithms admit
the $\mathcal{O}\left(N^{4}\right)$ complexity on constructing the
surrogate problems at each iteration.

\subsection{Convergence Analysis}

The convergence properties for the proposed algorithms are given in
the following.
\begin{prop}
\textcolor{black}{\label{prop: convergence of MM} Every limit point
of the solution sequence $\left\{ \mathbf{w}^{k}\right\} $ generated
by the Algorithm \ref{alg: MM for mvsk portfolio} is a stationary
point of problem (\ref{eq: mvsk portfolio}).}
\end{prop}
\begin{IEEEproof}
\textcolor{black}{Note that: 1) $\check{f}\left(\mathbf{w},\mathbf{w}^{k}\right)$
is continuous in both $\mathbf{w}$ and $\mathbf{w}^{k}$; 2) $\check{f}\left(\mathbf{w},\mathbf{w}^{k}\right)$
is a global upper bound function for $f\left(\mathbf{w}\right)$ and
is tangent to it at $\mathbf{w}^{k}$. Thus, \cite[Assumption 1]{razaviyayn2013unified}
is satisfied, and the proof of Proposition \ref{prop: convergence of MM}
follows directly from \cite[Theorem 1]{razaviyayn2013unified}.}
\end{IEEEproof}
\textcolor{black}{}
\begin{prop}
\textcolor{black}{\label{prop: convergence of SCA for MVSK} Suppose
$\gamma^{k}\in(0,1]$, $\gamma^{k}\rightarrow0$ and $\sum_{k}\gamma^{k}=+\infty$,
and let $\left\{ \mathbf{w}^{k}\right\} $ be the sequence generated
by Algorithm \ref{alg: SCA for mvsk portfolio}. Then either Algorithm
\ref{alg: SCA for mvsk portfolio} converges in a finite number of
iterations to a stationary point of (\ref{eq: mvsk portfolio}) or
every limit of $\left\{ \mathbf{w}^{k}\right\} $ (at least one such
point exists) is a stationary point of (\ref{eq: mvsk portfolio}).}
\end{prop}
\textcolor{blue}{}
\begin{IEEEproof}
\textcolor{black}{Note that the surrogate problem in Algorithm \ref{alg: SCA for mvsk portfolio}
only approximates the objective of problem (\ref{eq: mvsk portfolio})
with a quadratic one but leave the constraints untouched, and: 1)
$\mathcal{W}$ is a compact and convex set; 2) $f\left(\mathbf{w}\right)$
is continuously differentiable and coercive on $\mathcal{W}$; 3)
$\triangledown f_{\mathbf{w}}$ is Lipschitz continuous on $\mathcal{W}$
(provided by Lemma \ref{lem: bound for hessian matrix MM alg.}).
 Thus, \cite[Assumptions A1-A4]{scutari2013decomposition} are satisfied,
and the proof of Proposition \ref{prop: convergence of SCA for MVSK}
follows directly from \cite[Theorem 3]{scutari2013decomposition}.}
\end{IEEEproof}
\textcolor{black}{}
\begin{prop}
\textcolor{black}{\label{prop: convergence of mvsk tilting}Suppose
$\gamma^{k}\in(0,1]$, $\gamma^{k}\rightarrow0$ and $\sum_{k}\gamma^{k}=+\infty$,
and let $\left\{ \mathbf{w}^{k}\right\} $ be the sequence generated
by Algorithm \ref{alg: SCA for mvsk tilting linecon} or Algorithm
\ref{alg: SCA for mvsk tilting quadcon}. Then $\left\{ \mathbf{w}^{k}\right\} $
is a generalized stationary point of the problem (\ref{eq: mvsk tilting smooth}).}
\end{prop}
\textcolor{blue}{}
\begin{IEEEproof}
\textcolor{black}{The only difference between Algorithm \ref{alg: SCA for mvsk tilting linecon}
and Algorithm \ref{alg: SCA for mvsk tilting quadcon} is that Algorithm
\ref{alg: SCA for mvsk tilting quadcon} constructs the quadratic
approximation for the nonconvex constraints while the Algorithm \ref{alg: SCA for mvsk tilting linecon}
simply linearize all the constraints. However, it does not affect
the convergence checking as they are both convex approximation for
the constraints. Besides, it is easy to check that all the conditions
in Assumption \ref{asm: convergence conditions for general SCA} are
satisfied in both algorithms. Then the proof of Proposition \ref{prop: convergence of mvsk tilting}
follows directly from \cite{facchinei2020ghost}.}
\end{IEEEproof}

\section{Solving Other High-order Portfolio Problems \label{sec: other formulations}}

The algorithm framework proposed in this paper can be easily employed
to solve other high-order portfolio problems. 

\subsection{MVSK Tilting Portfolio with General Deterioration Measures}

As in \cite{BOUDT2020e03516}, the MVSK tilting portfolio problem
with general difference constraint to the reference portfolio is given
as follows:
\begin{equation}
\begin{aligned}\underset{\mathbf{w},\delta}{\mathsf{minimize}}\,\,\, & \quad-\delta\\
\mathsf{subject\,\,to} & \quad g_{\textrm{ref}}\left(\mathbf{w}\right)\le\kappa,\\
 & \quad g_{i}\left(\mathbf{w},\delta\right)\le0,\,\,i=1,\dots,4,\\
 & \quad\mathbf{w}\in\mathcal{W},\delta\ge0,
\end{aligned}
\label{eq: MVST tilting with general cons}
\end{equation}
where $g_{\textrm{ref}}\left(\mathbf{w}\right)$ is a measure of how
distant the current portfolio is from the reference one and $\kappa$
determines the maximum distance. For examples, $g_{\textrm{ref}}\left(\mathbf{w}\right)$
may be chosen as the risk concentration \cite{feng2015scrip}:
\begin{equation}
g_{\text{ref}}\left(\mathbf{w}\right)=\sum_{i=1}^{N}\left(\frac{w_{i}\left(\boldsymbol{\Sigma}\mathbf{w}\right)_{i}}{\mathbf{w}^{T}\boldsymbol{\Sigma}\mathbf{w}}-\frac{1}{N}\right)^{2}.
\end{equation}
The regularized MVSK tilting portfolio problem is obtained by transforming
the general distance constraint of problem (\ref{eq: MVST tilting with general cons})
to a regularization term in the objective:
\begin{equation}
\begin{aligned}\underset{\mathbf{w},\delta}{\mathsf{minimize}}\,\,\, & \quad-\delta+\lambda g_{\textrm{ref}}\left(\mathbf{w}\right)\\
\mathsf{subject\,\,to} & \quad g_{i}\left(\mathbf{w},\delta\right)\le0,\,\,i=1,\dots,4,\\
 & \quad\mathbf{w}\in\mathcal{W},\delta\ge0.
\end{aligned}
\label{eq: reg-MVST tilting with general cons}
\end{equation}
Obviously, problems \ref{eq: MVST tilting with general cons} and
\ref{eq: reg-MVST tilting with general cons} are both solvable via
the proposed algorithm framework in Section \ref{sec: solving MVSK tilting}.
The only difference is that here we also need to construct the convex
approximating function for $g_{\textrm{ref}}\left(\mathbf{w}\right)$
if it is nonconvex. The procedure is trivial and hence omitted.

\subsection{General Minkovski Distance MVST Portfolio}

The general Minkovski distance MVST portfolio \cite{nijkamp1980interactive}
admits the formulation
\begin{equation}
\begin{aligned}\underset{\mathbf{w},\mathbf{d}}{\mathsf{minimize}}\,\,\, & \quad z\left(\mathbf{d}\right)=\left(\sum_{k=1}^{4}\bigg\vert\frac{d_{k}}{z_{k}}\bigg\vert^{p}\right)^{1/p}\\
\mathsf{subject\,\,to} & \quad y_{i}\left(\mathbf{w},\mathbf{d}\right)\le0,i=1,\dots,4,\\
 & \quad\mathbf{w}\in\mathcal{W},\mathbf{d}\ge\mathbf{0}.
\end{aligned}
\end{equation}
where $z_{k}$ is the aspired levels for $k$-th moments and 
\begin{equation}
\begin{aligned}y_{1}\left(\mathbf{w},\mathbf{d}\right) & =-\phi_{1}\left(\mathbf{w}\right)-d_{1}+z_{1},\\
y_{2}\left(\mathbf{w},\mathbf{d}\right) & =\phi_{2}\left(\mathbf{w}\right)-d_{2}-z_{2},\\
y_{3}\left(\mathbf{w},\mathbf{d}\right) & =-\phi_{3}\left(\mathbf{w}\right)-d_{3}+z_{3},\\
y_{4}\left(\mathbf{w},\mathbf{d}\right) & =\phi_{4}\left(\mathbf{w}\right)-d_{4}-z_{4}.
\end{aligned}
\end{equation}
It is easy to write a sequence of convex approximating surrogate
problem as
\[
\begin{aligned}\underset{\mathbf{w},\mathbf{d}}{\mathsf{minimize}}\,\,\, & \quad\triangledown z(\mathbf{d}^{k})^{T}(\mathbf{d}-\mathbf{d}^{k})+\frac{\tau_{\mathbf{d}}}{2}\Vert\mathbf{d}-\mathbf{d}^{k}\Vert_{2}^{2}\\
 & \quad+\frac{\tau_{\mathbf{w}}}{2}\Vert\mathbf{w}-\mathbf{w}^{k}\Vert_{2}^{2}\\
\mathsf{subject\,\,to} & \quad\tilde{y}_{i}\left(\mathbf{w},\mathbf{d};\mathbf{w}^{k},\mathbf{d}^{k}\right)\le\eta(\mathbf{w}^{k},\mathbf{d}^{k}),i=1,\dots,4,\\
 & \quad\mathbf{w}\in\mathcal{W},\mathbf{d}\ge\mathbf{0}.
\end{aligned}
\]
where $\tilde{y}_{i}\left(\mathbf{w},\mathbf{d};\mathbf{w}^{k},\mathbf{d}^{k}\right)$
is the convex approximation of $y_{j}\left(\mathbf{w},\mathbf{d}\right)$
at $\left(\mathbf{w}^{k},\mathbf{d}^{k}\right)$, which can be easily
constructed following the similar procedures in Section \ref{sec: solving MVSK tilting}.

\subsection{Polynomial Goal Programming MVST Portfolio}

The polynomial goal programming (PGP) model for solving the high-order
portfolio \cite{lai2006mean,aksarayli2018polynomial} is a variation
of the general Minkovski distance MVST portfolio taking investors'
relative preference into consideration. It is formulated as
\begin{equation}
\begin{aligned}\underset{\mathbf{w},\mathbf{d}}{\mathsf{minimize}}\,\,\, & \quad z\left(\mathbf{d}\right)=\bigg\vert\frac{d_{1}}{z_{1}}\bigg\vert^{\lambda_{1}}+\bigg\vert\frac{d_{2}}{z_{2}}\bigg\vert^{\lambda_{2}}+\bigg\vert\frac{d_{3}}{z_{3}}\bigg\vert^{\lambda_{3}}+\bigg\vert\frac{d_{4}}{z_{4}}\bigg\vert^{\lambda_{4}}\\
\mathsf{subject\,\,to} & \quad y_{i}\left(\mathbf{w},\mathbf{d}\right)\le0,i=1,\dots,4,\\
 & \quad\mathbf{w}\in\mathcal{W},\mathbf{d}\ge\mathbf{0}.
\end{aligned}
\end{equation}
This problem can still be easily handled via the similar procedure
in solving the general Minkovski distance MVST portfolio.

\section{Numerical Experiments \label{sec: Numerical Experiments}}

In this section, we perform the numerical experiments on our proposed
algorithms \footnote{\textcolor{black}{We have released an R package $\mathsf{highOrderPortfolios}$
implementing our proposed algorithms at \href{https://github.com/dppalomar/highOrderPortfolios}{https://github.com/dppalomar/highOrderPortfolios}.}}. The data is generated according to the following steps: 
\begin{enumerate}
\item randomly select $N$ stocks from a dataset of 500 stocks, each of
them listed in the S\&P 500 Index components;
\item randomly pick $5N$ continuous trading days from 2004-01-01 to 2018-12-31;
\item compute four sample moments of the selected $N$ stocks during the
picked trading period.
\end{enumerate}
The starting point are selected as $\mathbf{w}^{0}=\frac{1}{N}\mathbf{1}$
for all methods. Without loss of generality, we simply set $L=1$,
$\theta=\frac{1}{2}$, and choose the diminishing step size sequence
as: 
\begin{equation}
\gamma^{0}=1,\quad\gamma^{k}=\gamma^{k-1}\left(1-10^{-2}\gamma^{k}\right).
\end{equation}
The inner solvers for QP, LP, and QCQP are selected as $\mathsf{quadprog}$
\cite{quadprog}, $\mathsf{lpSolveAPI}$ \cite{lpSolveAPI}, and $\mathsf{ECOS}$
\cite{domahidi2013ecos,ECOSolveR}, respectively. The algorithm is
regarded as converged when any of the following condition is satisfied:
\begin{equation}
\begin{aligned}\vert\mathbf{x}^{k+1}-\mathbf{x}^{k}\vert & \le10^{-6}\left(\vert\mathbf{x}^{k+1}\vert+\vert\mathbf{x}^{k}\vert\right),\\
\vert f(\mathbf{x}^{k+1})-f(\mathbf{x}^{k})\vert & \le10^{-6}\left(\vert f(\mathbf{x}^{k+1})\vert+\vert f(\mathbf{x}^{k})\vert\right).
\end{aligned}
\end{equation}

\subsection{On the MVSK Portfolio Problem (\ref{eq: mvsk portfolio}) \label{subsec: num res for mvsk solving}}

\begin{figure}
\begin{centering}
\includegraphics[scale=0.5]{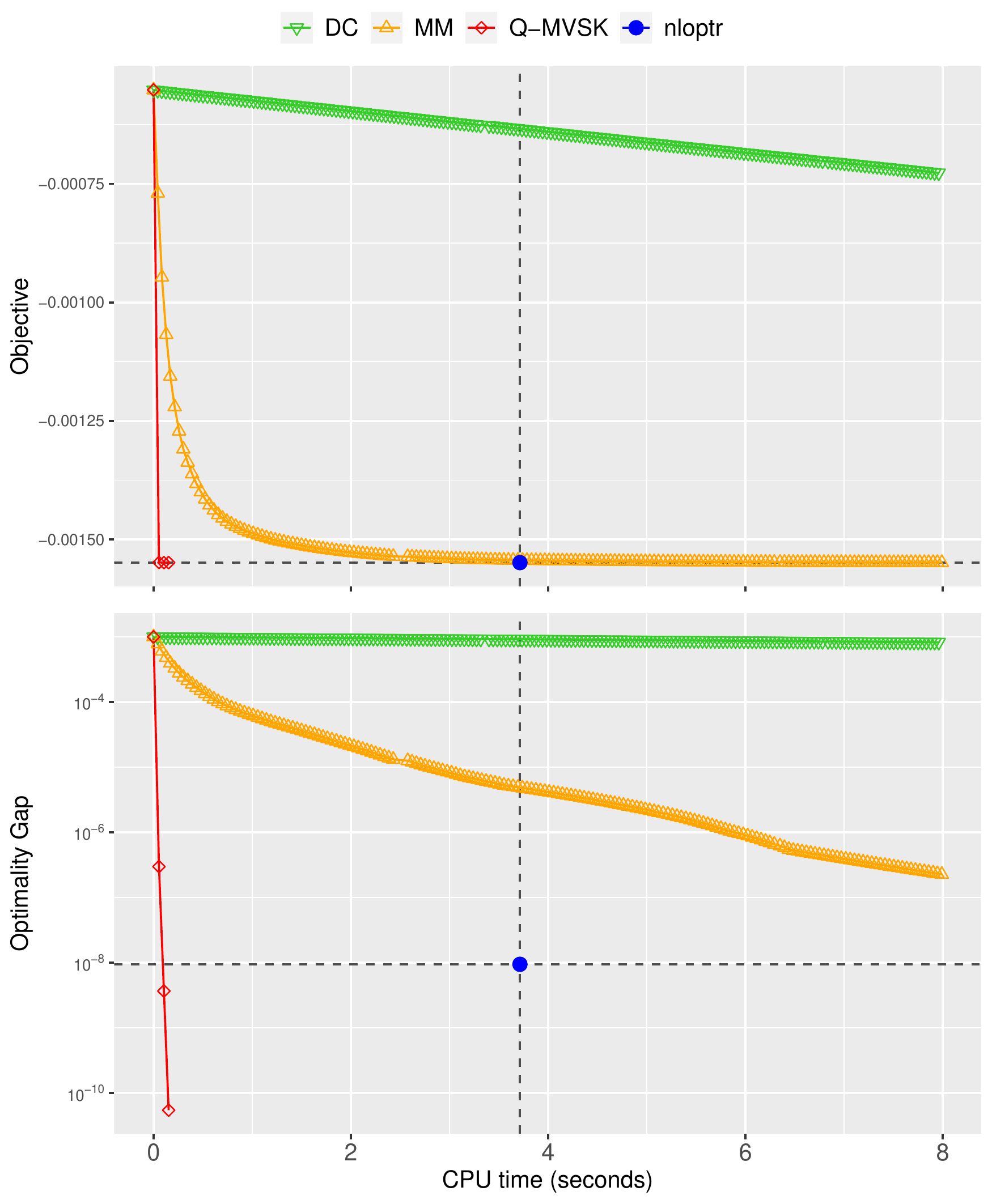}
\par\end{centering}
\caption{The convergence of algorithms on solving MVSK problem (\ref{eq: mvsk portfolio})
with $N=100$. \label{fig: convergence on mvsk}}
\end{figure}

We first set $N=100$ and then solve the problem (\ref{eq: mvsk portfolio})
using the DC-based Algorithm \ref{alg: DC for mvsk portfolio}, our
proposed MM-based Algorithm \ref{alg: MM for mvsk portfolio}, and
the Q-MVSK Algorithm \ref{alg: SCA for mvsk portfolio}. The weights
for four moments are decided according to the fourth order expansion
of the Constant Relative Risk Aversion (CRRA) utility function: 
\begin{equation}
\begin{aligned}\lambda_{1} & =1, & \lambda_{2} & =\frac{\xi}{2},\\
\lambda_{3} & =\frac{\xi\left(\xi+1\right)}{6}, & \lambda_{4} & =\frac{\xi\left(\xi+1\right)\left(\xi+2\right)}{24},
\end{aligned}
\end{equation}
where $\xi\ge0$ is the risk aversion parameter \cite{boudt2015higher}
and set to be $10$ in our experiments. For comparison, we also solve
the problem using the general optimization tool $\mathsf{nloptr}$
\cite{nloptr} with gradients passed. In Figure \ref{fig: convergence on mvsk},
we compare the convergence of these algorithms. Significantly, the
Q-MVSK algorithm can converge to the best result in very few iterations,
which is much more efficient than the solver $\mathsf{nloptr}$. The
DC-based and MM-based algorithms are both slower than the general
solver $\mathsf{nloptr}$. It implies that they may use very loose
upper bounds. The MM-based algorithm, though much faster than the
DC-based algorithm, is far from being comparable with the Q-MVSK algorithm.

\begin{figure}
\begin{centering}
\includegraphics[scale=0.58]{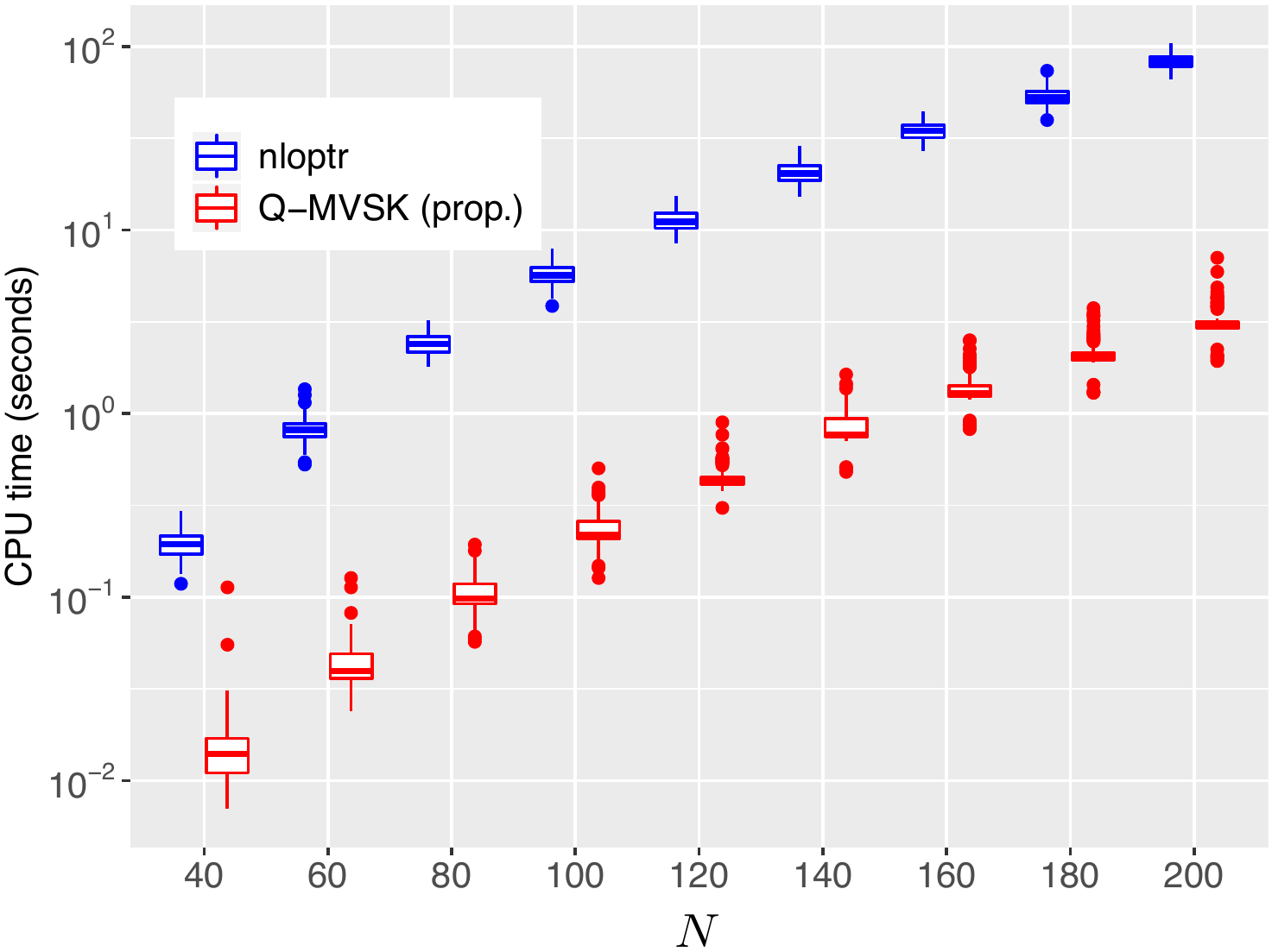}
\par\end{centering}
\caption{Time usage of algorithms on solving MVSK problem (\ref{eq: mvsk portfolio}).
\label{fig: time on mvsk}}
\end{figure}

In Figure \ref{fig: time on mvsk}, we show the comparison of time
consumption of the proposed Q-MVSK algorithm and $\mathsf{nloptr}$
while changing the problem dimension $N$. \textcolor{black}{The DC-based
and the MM-based algorithms are not included as they are too slow
to be compared with the proposed Q-MVSK algorithm and $\mathsf{nloptr}$.}
For fair comparison, we force $\mathsf{nloptr}$ to run until it reaches
the objective obtained from Q-MVSK algorithm. The result is obtained
by performing the experiments on $100$ realizations of randomly generated
data. We can see that our proposed Q-MVSK algorithm is consistently
more than one order of magnitude faster than the $\mathsf{nloptr}$.


\subsection{On the MVSK Tilting Portfolio Problem (\ref{eq: mvsk tilting problem})}

\begin{figure}
\begin{centering}
\includegraphics[scale=0.54]{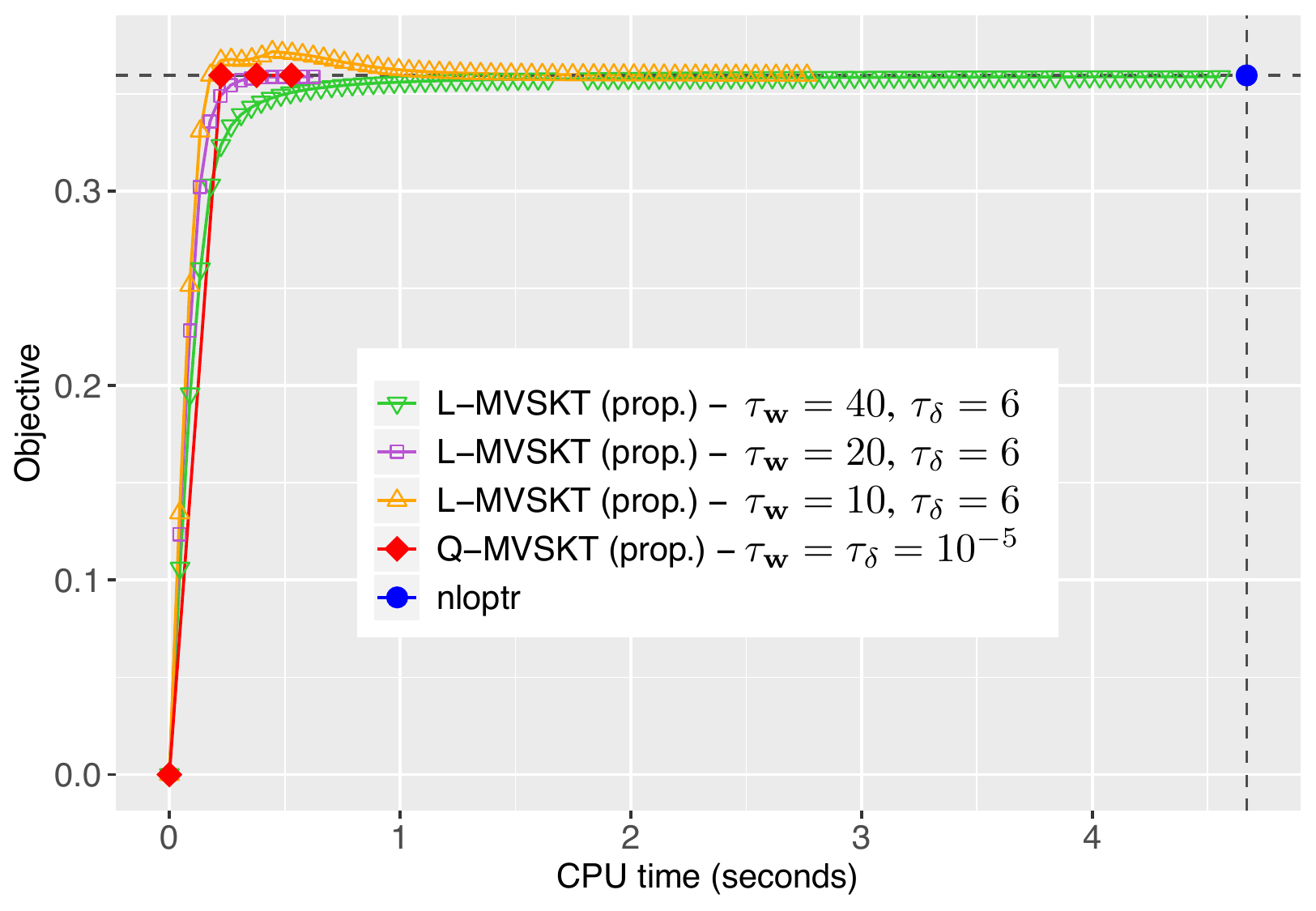}
\par\end{centering}
\caption{Convergence of proposed algorithms for MVSK tilting problem (\ref{eq: mvsk tilting problem})
with $N=100$ and $\kappa=0.3\sqrt{\phi_{2}(\mathbf{w}_{0})}$. \label{fig: convergence of mvsk tilting c=00003D0.3}}
\end{figure}

\begin{figure}
\begin{centering}
\includegraphics[scale=0.54]{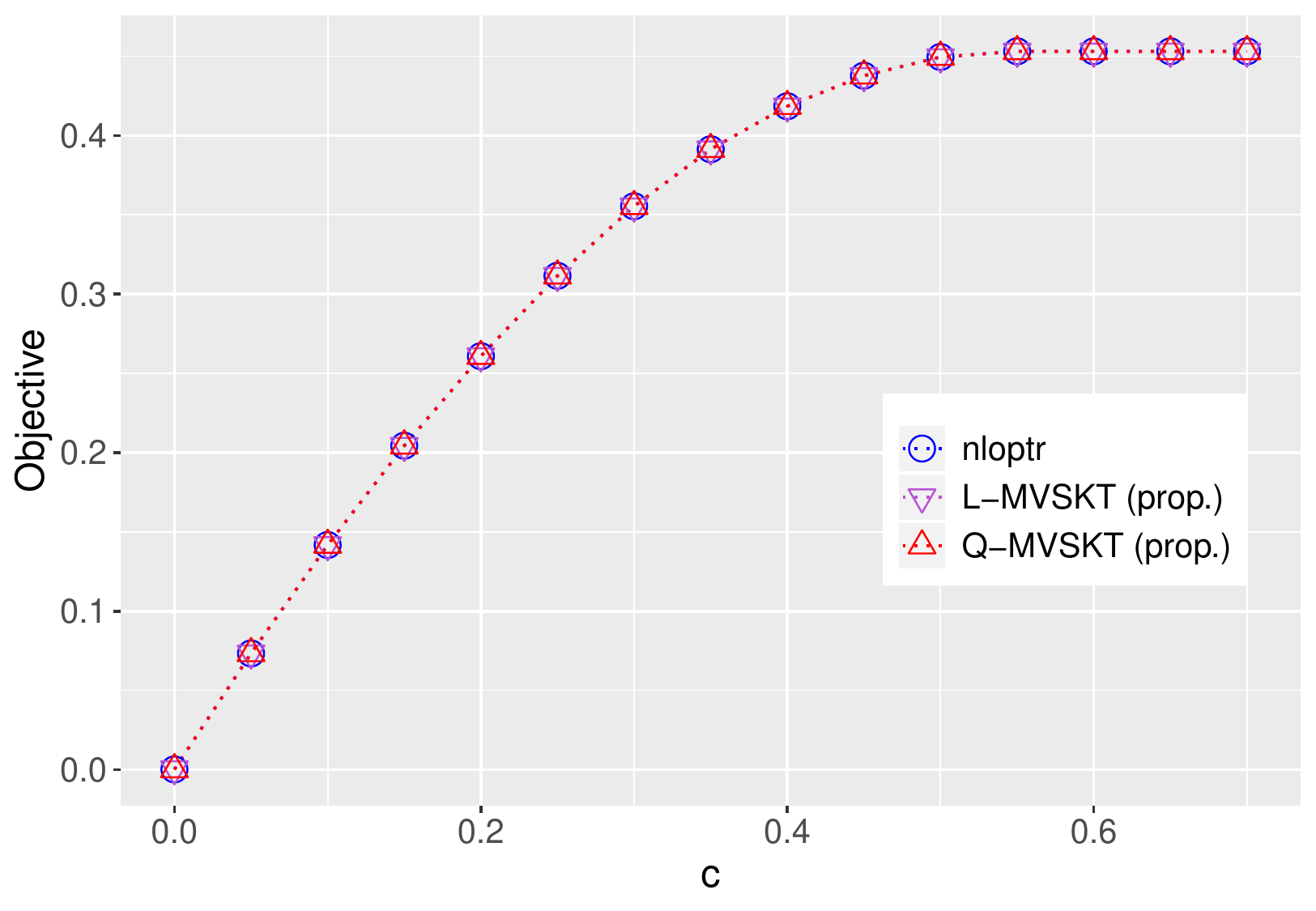}
\par\end{centering}
\caption{Comparison of the results with different $\kappa$ ($\kappa=c\times\sqrt{\phi_{2}(\mathbf{w}_{0})}$).
\label{fig: obj of mvsk tilting with diff kappa}}
\end{figure}

Similar to the above, we first set $N=100$ and then solve the problem
(\ref{eq: mvsk tilting problem}) via the proposed Algorithms \ref{alg: SCA for mvsk tilting linecon}
and \ref{alg: SCA for mvsk tilting quadcon}, respectively. The reference
portfolio is simply chosen as the equally weighted portfolio, i.e.,
\begin{equation}
\mathbf{w}_{0}=\frac{1}{N}\mathbf{1}.
\end{equation}
\textcolor{black}{The tilting direction $\mathbf{d}$ is decided as
$d_{i}=\vert\phi_{i}(\mathbf{w}_{0})\vert$.} We choose $\kappa$
in (\ref{eq: mvsk tilting problem}) as $\kappa=c\times\sqrt{\phi_{2}(\mathbf{w}_{0})}$
with $c\ge0$. The general solver $\mathsf{nloptr}$ is also included
for comparison \footnote{\textcolor{black}{We use directly the implementation from authors
of \cite{BOUDT2020e03516}, which is available at \href{https://github.com/cdries/mvskPortfolios}{https://github.com/cdries/mvskPortfolios}.}}. We find that, although the final convergence is guaranteed, the
fast convergence of the proposed L-MVSKT algorithm really relies on
the proper choice of $\tau_{\mathbf{w}}$ and $\tau_{\delta}$, while
that of our proposed Q-MVSKT is much robust. For example, in Figure
\ref{fig: convergence of mvsk tilting c=00003D0.3}, we set $\kappa=0.3\sqrt{\phi_{2}(\mathbf{w}_{0})}$
and show the convergence of the proposed algorithms.  It is significant
that the Q-MVSKT algorithm converges in few iterations simply with
$\tau_{\mathbf{w}}=\tau_{\delta}=10^{-5}$. The L-MVSKT algorithm
can also converge with comparable speed when parameters are properly
tuned. It may be explained as that the L-MVSKT algorithm poorly approximates
all constraints by linear functions, making the solution to approximating
problems easily violates the original constraints. However, the Q-MVSKT
algorithm preserves the convex constraints and approximates the nonconvex
constraints by convex quadratic functions, which turns out to work
very well. Besides, we notice that solving the QCQP problem is significantly
slower than solving the QP problem of the same size. It might be because
we are using the R interface to a more general second-order cone programming
(SOCP) solver, i.e., $\mathsf{ECOS}$ \cite{ECOSolveR}. In Figure
\ref{fig: obj of mvsk tilting with diff kappa}, we show the final
results of these algorithms when changing the maximum tracking error
constraint. It is clear that all algorithms can give the same results,
which are nondecreasing when $\kappa$ increases. 
\begin{figure}
\begin{centering}
\includegraphics[scale=0.58]{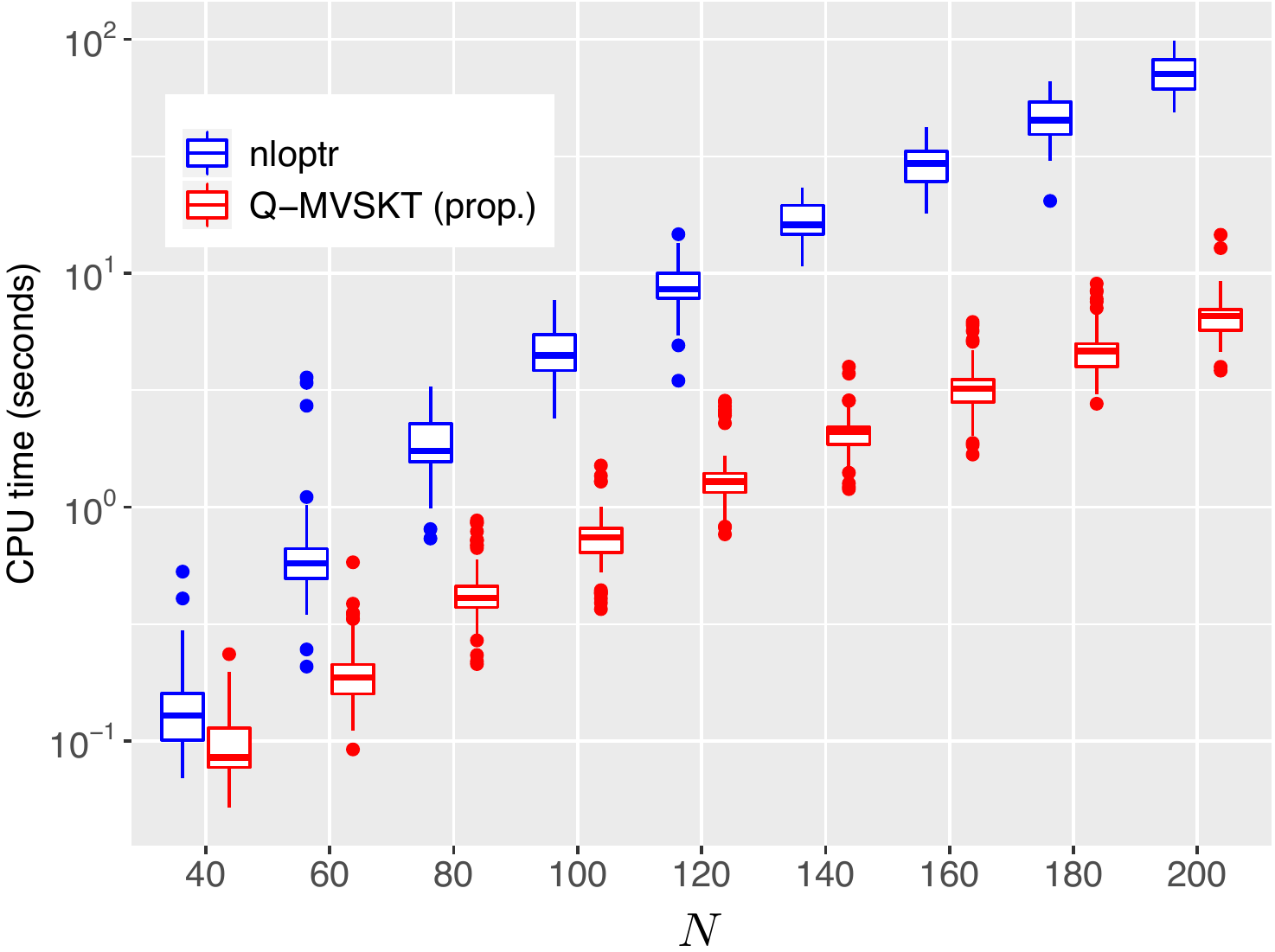}
\par\end{centering}
\caption{Time usage of algorithms on solving problem (\ref{eq: mvsk tilting problem}).
\label{fig: time of mvsk tilting}}
\end{figure}

In Figure \ref{fig: time of mvsk tilting}, we show the comparison
of time consumption of the proposed Q-MVSKT algorithm and $\mathsf{nloptr}$
while changing the problem dimension $N$. \textcolor{black}{The proposed
L-MVSKT algorithm is not included as its convergence speed relies
heavily on parameter tuning.} The result is obtained by performing
the experiments on $100$ realizations of randomly generated data.
It is significant that the proposed Q-MVSKT consistently outperform
the L-MVSKT algorithm and is about one order of magnitude faster than
$\mathsf{nloptr}$.

\section{Conclusion \label{sec: conclusion}}

In this paper, we have considered the high-order moments of the portfolio
return for high-order portfolio optimization. We have proposed an
efficient algorithm framework for solving the high-order portfolio
optimization problems based on the successive convex approximation
framework. In particular, we have proposed efficient algorithms for
solving the mean-variance-skewness-kurtosis portfolio optimization
problem and the mean-variance-skewness-kurtosis tilting portfolio
optimization problem. Theoretically, all the proposed algorithms enjoy
global convergence to a stationary point. Extensive numerical experiments
show that our proposed algorithms, \textcolor{black}{specifically
the Q-MVSK and Q-MVSKT algorithms,} are much more efficient than the
existing method and the general solver.

\appendix{}

\subsection{Proof for Lemma \ref{lem: gradient and hessian of moments} \label{apx: proof grad and hsn}}

According to the Leibniz integral rule \cite{abramowitz1948handbook},
we have

\begin{equation}
\begin{aligned}\triangledown\phi_{3}\left(\mathbf{w}\right) & =\frac{\partial\text{E}\left[\mathbf{w}^{T}\tilde{\mathbf{r}}\tilde{\mathbf{r}}^{T}\mathbf{w}\tilde{\mathbf{r}}^{T}\mathbf{w}\right]}{\partial\mathbf{w}}\\
 & =\text{E}\left[\frac{\partial\left(\mathbf{w}^{T}\tilde{\mathbf{r}}\mathbf{w}^{T}\tilde{\mathbf{r}}\tilde{\mathbf{r}}^{T}\mathbf{w}\right)}{\partial\mathbf{w}}\right]\\
 & =\text{E}\left[3\tilde{\mathbf{r}}\left(\tilde{\mathbf{r}}^{T}\otimes\tilde{\mathbf{r}}^{T}\right)\left(\mathbf{w}\otimes\mathbf{w}\right)\right]\\
 & =3\text{E}\left[\tilde{\mathbf{r}}\left(\tilde{\mathbf{r}}^{T}\otimes\tilde{\mathbf{r}}^{T}\right)\right]\left(\mathbf{w}\otimes\mathbf{w}\right)\\
 & =3\boldsymbol{\Phi}\left(\mathbf{w}\otimes\mathbf{w}\right),
\end{aligned}
\end{equation}
\begin{equation}
\begin{aligned}\triangledown\phi_{4}\left(\mathbf{w}\right) & =\frac{\partial\text{E}\left[\mathbf{w}^{T}\tilde{\mathbf{r}}\tilde{\mathbf{r}}^{T}\mathbf{w}\tilde{\mathbf{r}}^{T}\mathbf{w}\tilde{\mathbf{r}}^{T}\mathbf{w}\right]}{\partial\mathbf{w}}\\
 & =\text{E}\left[\frac{\partial\left(\mathbf{w}^{T}\tilde{\mathbf{r}}\tilde{\mathbf{r}}^{T}\mathbf{w}\tilde{\mathbf{r}}^{T}\mathbf{w}\tilde{\mathbf{r}}^{T}\mathbf{w}\right)}{\partial\mathbf{w}}\right]\\
 & =\text{E}\left[4\tilde{\mathbf{r}}\left(\tilde{\mathbf{r}}^{T}\otimes\tilde{\mathbf{r}}^{T}\otimes\tilde{\mathbf{r}}^{T}\right)\left(\mathbf{w}\otimes\mathbf{w}\otimes\mathbf{w}\right)\right]\\
 & =4\text{E}\left[\tilde{\mathbf{r}}\left(\tilde{\mathbf{r}}^{T}\otimes\tilde{\mathbf{r}}^{T}\otimes\tilde{\mathbf{r}}^{T}\right)\right]\left(\mathbf{w}\otimes\mathbf{w}\otimes\mathbf{w}\right)\\
 & =4\boldsymbol{\Psi}\left(\mathbf{w}\otimes\mathbf{w}\otimes\mathbf{w}\right),
\end{aligned}
\end{equation}
\begin{equation}
\begin{aligned}\triangledown^{2}\phi_{3}\left(\mathbf{w}\right) & =\frac{\partial^{2}\text{E}\left[\mathbf{w}^{T}\tilde{\mathbf{r}}\tilde{\mathbf{r}}^{T}\mathbf{w}\tilde{\mathbf{r}}^{T}\mathbf{w}\right]}{\partial\mathbf{w}\partial\mathbf{w}^{T}}\\
 & =\text{E}\left[\frac{\partial^{2}\left(\mathbf{w}^{T}\tilde{\mathbf{r}}\mathbf{w}^{T}\tilde{\mathbf{r}}\tilde{\mathbf{r}}^{T}\mathbf{w}\right)}{\partial\mathbf{w}\partial\mathbf{w}^{T}}\right]\\
 & =\text{E}\left[6\tilde{\mathbf{r}}\left(\tilde{\mathbf{r}}^{T}\otimes\tilde{\mathbf{r}}^{T}\right)\left(\mathbf{I}\otimes\mathbf{w}\right)\right]\\
 & =6\text{E}\left[\tilde{\mathbf{r}}\left(\tilde{\mathbf{r}}^{T}\otimes\tilde{\mathbf{r}}^{T}\right)\right]\left(\mathbf{I}\otimes\mathbf{w}\right)\\
 & =6\boldsymbol{\Phi}\left(\mathbf{I}\otimes\mathbf{w}\right),
\end{aligned}
\end{equation}
\begin{equation}
\begin{aligned}\triangledown^{2}\phi_{4}\left(\mathbf{w}\right) & =\frac{\partial^{2}\text{E}\left[\mathbf{w}^{T}\tilde{\mathbf{r}}\tilde{\mathbf{r}}^{T}\mathbf{w}\tilde{\mathbf{r}}^{T}\mathbf{w}\tilde{\mathbf{r}}^{T}\mathbf{w}\right]}{\partial\mathbf{w}\partial\mathbf{w}^{T}}\\
 & =\text{E}\left[\frac{\partial^{2}\left(\mathbf{w}^{T}\tilde{\mathbf{r}}\tilde{\mathbf{r}}^{T}\mathbf{w}\tilde{\mathbf{r}}^{T}\mathbf{w}\tilde{\mathbf{r}}^{T}\mathbf{w}\right)}{\partial\mathbf{w}\partial\mathbf{w}^{T}}\right]\\
 & =\text{E}\left[12\tilde{\mathbf{r}}\left(\tilde{\mathbf{r}}^{T}\otimes\tilde{\mathbf{r}}^{T}\otimes\tilde{\mathbf{r}}^{T}\right)\left(\mathbf{I}\otimes\mathbf{w}\otimes\mathbf{w}\right)\right]\\
 & =12\text{E}\left[\tilde{\mathbf{r}}\left(\tilde{\mathbf{r}}^{T}\otimes\tilde{\mathbf{r}}^{T}\otimes\tilde{\mathbf{r}}^{T}\right)\right]\left(\mathbf{I}\otimes\mathbf{w}\otimes\mathbf{w}\right)\\
 & =12\boldsymbol{\Psi}\left(\mathbf{I}\otimes\mathbf{w}\otimes\mathbf{w}\right).
\end{aligned}
\end{equation}

\subsection{Proof for Lemma \ref{lem: bound for hessian matrix MM alg.} \label{apx: proof MM hsn bound}}

According to the Gershgorin circle theorem \cite{varga2010gervsgorin},
we have
\begin{equation}
\begin{aligned} & \rho\left(\triangledown^{2}f_{\text{ncvx}}\left(\mathbf{w}\right)\right)\\
 & \le\Vert\triangledown^{2}f_{\text{ncvx}}\left(\mathbf{w}\right)\Vert_{\infty}\\
 & \le\lambda_{3}\Vert\triangledown^{2}\phi_{3}\left(\mathbf{w}\right)\Vert_{\infty}+\lambda_{4}\Vert\triangledown^{2}\phi_{4}\left(\mathbf{w}\right)\Vert_{\infty}.
\end{aligned}
\end{equation}
Under the constraints in (\ref{eq: constrains for w}), we can get
\begin{equation}
\begin{aligned} & \Vert\triangledown^{2}\phi_{3}\left(\mathbf{w}\right)\Vert_{\infty}\\
 & =6\max_{1\le i\le N}\sum_{j=1}^{N}\lvert\sum_{k=1}^{N}\Phi_{ij}^{(k)}w_{k}\rvert\\
 & \le6\max_{1\le i\le N}\sum_{j=1}^{N}\sum_{k=1}^{N}\lvert\Phi_{ij}^{(k)}\lvert\lvert w_{k}\rvert\\
 & \le6\max_{1\le i\le N}\sum_{j=1}^{N}\max_{1\le k\le N}L\vert\Phi_{ij}^{(k)}\vert\\
 & =6L\max_{1\le i\le N}\sum_{j=1}^{N}\max_{1\le k\le N}\vert\Phi_{ij}^{(k)}\vert,
\end{aligned}
\end{equation}
\begin{equation}
\begin{aligned} & \Vert\triangledown^{2}\phi_{4}\left(\mathbf{w}\right)\Vert_{\infty}\\
 & =12\max_{1\le i\le N}\sum_{j=1}^{N}\lvert\sum_{k=1}^{N}w_{k}\sum_{l=1}^{N}\Psi_{ij}^{(k,l)}w_{l}\rvert\\
 & \le12\max_{1\le i\le N}\sum_{j=1}^{N}\sum_{k=1}^{N}\vert w_{k}\vert\sum_{l=1}^{N}\vert\Psi_{ij}^{(k,l)}\vert\vert w_{l}\vert\\
 & \le12\max_{1\le i\le N}\sum_{j=1}^{N}\sum_{k=1}^{N}\vert w_{k}\vert\max_{1\le l\le N}L\vert\Psi_{ij}^{(k,l)}\vert,\\
 & \le12\max_{1\le i\le N}\sum_{j=1}^{N}\max_{1\le k\le N}L\max_{1\le l\le N}L\vert\Psi_{ij}^{(k,l)}\vert\\
 & =12L^{2}\max_{1\le i\le N}\sum_{j=1}^{N}\max_{1\le k,l\le N}\vert\Psi_{ij}^{(k,l)}\vert.
\end{aligned}
\end{equation}
Therefore, we have 
\begin{equation}
\begin{aligned}\rho\left(\triangledown^{2}f_{\text{ncvx}}\left(\mathbf{w}\right)\right) & \le6\lambda_{3}L\max_{1\le i\le N}\sum_{j=1}^{N}\max_{1\le k\le N}\vert\Phi_{ij}^{(k)}\vert\\
 & \quad+12\lambda_{4}L^{2}\max_{1\le i\le N}\sum_{j=1}^{N}\max_{1\le k,l\le N}\vert\Psi_{ij}^{(k,l)}\vert.
\end{aligned}
\end{equation}

\bibliographystyle{IEEEtran}
\bibliography{refs}

\end{document}